\documentclass[11pt,a4paper,twocolumn]{article}

\usepackage[utf8]{inputenc}
\usepackage[T1]{fontenc}
\usepackage{lmodern}
\usepackage{amsmath,amssymb}
\usepackage{graphicx}
\usepackage{booktabs}
\usepackage{xcolor}
\usepackage{enumitem}
\usepackage{array}
\usepackage{tabularx}
\usepackage{float}
\usepackage[margin=0.75in]{geometry}
\usepackage{caption}
\usepackage{tikz}
\usepackage{pgfplots}
\usepackage{textcomp}
\usepackage{url}
\usepackage{hyperref}
\usepackage[protrusion=true,expansion=true,final]{microtype}
\usepackage{multirow}
\usepackage{longtable}

\usetikzlibrary{shapes.geometric, arrows.meta, positioning, calc, fit, backgrounds, patterns}
\pgfplotsset{compat=1.18}

\hypersetup{colorlinks=true, linkcolor=blue!60!black, citecolor=blue!60!black, urlcolor=blue!60!black}

\makeatletter
\renewenvironment{abstract}{%
  \begin{center}{\bfseries\large \abstractname\vspace{-0.5em}\vspace{\z@}}\end{center}%
  \quotation\small}%
  {\endquotation}
\makeatother

\definecolor{gapcolor}{HTML}{C0392B}
\definecolor{holdcolor}{HTML}{1E8449}
\definecolor{warncolor}{HTML}{D35400}
\definecolor{infocolor}{HTML}{2471A3}
\definecolor{platcolor}{HTML}{8E44AD}
\definecolor{manifestcolor}{HTML}{2980B9}
\definecolor{codecolor}{HTML}{27AE60}
\definecolor{archcolor}{HTML}{C0392B}
\definecolor{barblue}{HTML}{2471A3}
\definecolor{barred}{HTML}{C0392B}
\definecolor{bargreen}{HTML}{1E8449}
\definecolor{barorange}{HTML}{D35400}
\definecolor{barpurple}{HTML}{8E44AD}

\title{\textbf{A Systematic Study of Security Enforcement\\at Android Inter-Component Boundaries}\\[3pt]
\large An Empirical Evaluation of First-Party Interfaces\\and Platform Security Guardrails on Android~17}

\author{Sandiyo Christan A\\Independent Researcher\\sandichrist6@gmail.com}

\date{October 2026}

\begin{document}

\twocolumn[
\maketitle
\begin{abstract}
Android's inter-process communication (IPC) architecture connects applications to each other and to the system through Activities, Services, BroadcastReceivers, ContentProviders, and custom URI scheme handlers. We ran an expected-versus-observed study of security enforcement at these boundaries across over 30 Google and AOSP packages on Android~17 (API~37) under a zero-permission attacker model, covering 120 evaluations: 99 at inter-component interfaces, 10 of platform security controls, and 11 positive controls. In 97 evaluations the observed behavior diverged from the expected enforcement; from these we derive a 14-category taxonomy across manifest, code, platform, and architectural layers. Missing authorization on exported components and unvalidated ContentProvider query parameters are the most common implementation-level patterns. We break the 97 divergences down by evidence and kind: 54 are gaps confirmed dynamically (53 in exported Activities, ContentProviders, BroadcastReceivers, and Services; one a platform Parcelable regression); 3 confirm a missing check whose impact we could not establish; 27 are hardening or by-design observations, including 18 custom URI scheme registrations that carry sensitive data even though the platform cannot verify custom schemes (RFC~8252 \S8.1) and compatibility changes with security intent that are disabled or deferred by \texttt{targetSdkVersion} gating; 7 rest on static analysis only; 5 are crash-only availability issues; and one documents another researcher's CVE. We chose evaluations where gaps were expected, so we do not report an overall gap rate. The 23 evaluations in which enforcement held, eight in ContentProviders and four in Services, confirm that the procedure can tell working enforcement from failures, and they concentrate in the categories where the framework offers per-request enforcement primitives.
\end{abstract}

\begingroup
\vspace{0.6em}
\small\noindent\textbf{Keywords:}\enspace Android security, inter-process communication, IPC enforcement, ContentProvider, SQL injection, URI scheme, compatibility framework, zero-permission attacker, permission model, confused deputy\par
\endgroup
\vspace{1.6em}
]

\begin{table}[!tbp]
\centering \footnotesize
\caption{Abbreviations used in this paper.}
\label{tab:abbrev}
\begin{tabular}{@{}ll@{}}
\toprule
\textbf{Abbr.} & \textbf{Full Form} \\
\midrule
2FA & Two-Factor Authentication \\
ADB & Android Debug Bridge \\
AIDL & Android Interface Definition Language \\
AOSP & Android Open Source Project \\
API & Application Programming Interface \\
BAL & Background Activity Launch \\
CA & Certificate Authority \\
DEX & Dalvik Executable \\
DoS & Denial of Service \\
eSIM & Embedded SIM \\
FRP & Factory Reset Protection \\
GMS & Google Mobile Services \\
HCE & Host Card Emulation \\
ICC & Inter-Component Communication \\
IPC & Inter-Process Communication \\
NFC & Near Field Communication \\
PKCE & Proof Key for Code Exchange \\
PoC & Proof of Concept \\
SDK & Software Development Kit \\
SQL & Structured Query Language \\
TLPE & Third-party Library Proxy Execution \\
TOTP & Time-based One-Time Password \\
UID & User Identifier \\
UPI & Unified Payments Interface \\
URI & Uniform Resource Identifier \\
\bottomrule
\end{tabular}
\end{table}

\section{Introduction}
\label{sec:intro}

Android's inter-process communication (IPC) architecture lets applications expose functionality to other applications and to the system through Activities, Services, BroadcastReceivers, and ContentProviders. These four component types, together with custom URI scheme handlers, are the points where sensitive data and control cross between trust domains. Keeping those crossings secure requires enforcement at several levels at once: manifest-declared permissions, code-level caller identity checks, framework-provided input validation, and platform compatibility changes.

After over a decade of Android IPC security research, one basic question is still open: does enforcement at these boundaries actually hold? Prior work covers the problem in parts. Tools like ComDroid~\cite{chin2011}, CHEX~\cite{lu2012}, and IccTA~\cite{li2015} find vulnerability patterns automatically, at large scale, but they do not check enforcement as a whole. Published advisories record individual failures without tracing systemic patterns. Studies of how users understand the permission model~\cite{felt2012} and compositional security analyses~\cite{bagheri2015} examine the framework's design rather than how enforcement plays out at individual boundaries.

This study fills that gap with 120 security-enforcement evaluations across over 30 Google and AOSP packages on Android~17. Four things set our approach apart from the tools above. We frame each evaluation as a boundary assessment: does the enforcement mechanism at this boundary do what it is meant to do? We record both gaps (observed behavior differs from expected) and holds (enforcement works as designed), giving calibration cases that show the procedure can tell them apart. We cover five inter-component interface categories and platform security controls in one study, so cross-category comparison is possible. And we sort the divergences into confirmed-dynamic, static-only, crash-only, hardening, and by-design groups, so that an architectural limitation or a deliberate platform policy is not lumped in with an exploitable defect.

\paragraph{Research questions.}
\begin{enumerate}[label=\textbf{RQ\arabic*:}, leftmargin=2.6em, itemsep=1pt]
\item Across a purposively selected corpus of first-party applications, how does observed security enforcement differ among Android inter-component interface types?
\item What repeated implementation choices and structural properties account for the divergence between intended enforcement design and behavior observed in practice?
\item Do interface categories in which the framework supplies per-request enforcement primitives exhibit different observed outcomes from categories that depend mainly on application-level checks?
\end{enumerate}

\paragraph{Contributions.}
\begin{enumerate}[label=\arabic*., leftmargin=1.6em, itemsep=1pt]
\item \textbf{Expected-versus-observed framework.} An evaluation method that establishes what enforcement should occur at each boundary from documentation, sibling comparison, and platform design intent, then measures what enforcement actually occurs through proof-of-concept applications and code review (Section~\ref{sec:method}).

\item \textbf{120 evaluations with 23 positive controls.} Enforcement evaluations spanning Activities~(27), Content\-Providers~(29), Broadcast\-Receivers~(11), Services~(14), custom URI schemes~(18), and platform security controls~(10), with 11 separately evaluated well-defended components. Of the 120 evaluations, 23 confirm functioning enforcement and serve as calibration cases; of the 97 divergences, 54 carry dynamic confirmation and the rest are classified by kind (Section~\ref{sec:results}).

\item \textbf{14-category enforcement gap taxonomy.} An evidence-derived taxonomy that groups failure patterns across manifest, code, platform, and architectural layers (Section~\ref{sec:taxonomy}).

\item \textbf{Compatibility change deployment state.} A record of seven IPC-related compatibility changes with security intent observed on the test device: three are globally disabled and four activate only for applications targeting SDK~37. We treat them as staged-rollout decisions rather than defects, and discuss what they imply for a zero-permission attacker (Section~\ref{subsec:platform}).

\item \textbf{Cross-category enforcement comparison.} Evidence that enforcement outcomes vary substantially by interface category; categories that provide framework-level enforcement primitives show higher hold rates (Section~\ref{sec:discussion}).
\end{enumerate}

The remainder of the paper is organized as follows. Section~\ref{sec:background} covers Android IPC and the permission model. Methodology and threat model are in Section~\ref{sec:method}; findings in Section~\ref{sec:results}; the enforcement gap taxonomy in Section~\ref{sec:taxonomy}. Section~\ref{sec:discussion} discusses implications and validity. Section~\ref{sec:related} surveys related work, and Section~\ref{sec:conclusion} concludes.

\section{Background}
\label{sec:background}

\subsection{Android IPC Components}

Android applications expose capabilities through four component types, each with its own IPC behavior:

\textbf{Activities} are user-interface screens. Another app launches one by calling \texttt{startActivity()}, passing an Intent whose extras the target Activity reads. When the Activity is exported, any installed application may invoke it and supply attacker-controlled data in those extras.

\textbf{Services} handle background work. A calling app reaches a Service through \texttt{bindService()} or \texttt{startService()}. When bound, the Service may expose an AIDL interface that the caller can invoke as remote method calls.

\textbf{BroadcastReceivers} handle broadcast messages, whether sent system-wide or targeted at a specific package. The receiver reads data from the broadcast Intent, which can carry attacker-controlled content.

\textbf{ContentProviders} structure data access through a SQL-style interface (\texttt{query()}, \texttt{insert()}, \texttt{update()}, \texttt{delete()}), file retrieval via \texttt{openFile()}, and custom operations through \texttt{call()}. The caller provides a selection predicate, a column projection, and a sort order; the provider builds its database query from these parameters.

\subsection{Android Permission Model}

Android's permission system governs component access across four protection levels. \emph{Normal} permissions are granted automatically on install; \emph{dangerous} ones prompt the user for explicit approval. At the higher levels, \emph{signature} admits only callers whose signing certificate matches the defining app's certificate, while \emph{signature|privileged} further restricts access to platform-signed apps or apps on the privileged partition. A component declares its requirement in the manifest: Activities, Services, and BroadcastReceivers use the \texttt{android:\allowbreak{}permission} attribute; ContentProviders expose separate \texttt{android:\allowbreak{}read\-Permission} and \texttt{android:\allowbreak{}write\-Permission} attributes. A permission set on the \texttt{<application>} element acts as the default for every component in the app unless a specific component overrides it.

\subsection{Component Export and Intent Resolution}

A component becomes reachable outside its process either by explicitly setting \texttt{android:\allowbreak{}exported=\allowbreak{}"true"} or, for apps targeting API level~30 or lower, by declaring an intent filter (which triggers implicit export). The two attributes serve different roles: \texttt{android:\allowbreak{}exported} determines whether external callers can see and invoke the component, while \texttt{android:\allowbreak{}permission} determines whether they are authorized to do so. Exporting a component that carries no permission attribute, with no application-level default, opens it to any installed app without restriction.

\subsection{ContentProvider Query Defenses}
\label{subsec:cpdefenses}

Three opt-in defenses against ContentProvider query injection are available through \texttt{SQLiteQueryBuilder}, each requiring explicit developer activation. Calling \texttt{setStrict(true)} checks the selection for injected clauses; when combined with a projection map, it also rejects columns outside that map. \texttt{setStrictColumns(true)} covers projections and write values in the same way. \texttt{setStrictGrammar(true)} goes further, rejecting unexpected SQL keywords or subqueries in selections and sort orders. The compatibility change \texttt{ENFORCE\_\allowbreak{}STRICT\_\allowbreak{}QUERY\_\allowbreak{}BUILDER} (ChangeId~143231523) was introduced to activate \texttt{SQLiteQueryBuilder} validation for specific ContentProvider queries; its documented scope covers CalendarProvider2 query selections.

\subsection{Custom URI Schemes and App Links}

Custom URI schemes such as \texttt{otpauth://} and \texttt{gpay://} are registered by adding intent-filters to the manifest. HTTPS App Links can be tied to a verified domain through \texttt{android:\allowbreak{}autoVerify}, but no parallel verification exists for custom schemes. Because Android cannot determine which app legitimately owns a custom scheme, any installed app may declare itself a handler for it, creating an interception path. RFC~8252 \S8.1~\cite{rfc8252} identifies this architectural risk and recommends PKCE for public native clients, so that an authorization code captured in transit cannot be exchanged by an intercepting app.

\subsection{Platform Compatibility Framework}

Android's Compatibility Framework gates behavioral changes on the target application's \texttt{targetSdkVersion}. A change may be globally on, globally off, or on only for applications that meet a minimum SDK target. The staged rollout is deliberate: existing applications keep working until their developers raise the target and adapt. The security consequence is narrow but real. A hardening change covers an application only after that application has raised its target to the gating level. An application that has not updated, or an attacker application that declares a lower target, falls outside the change's protection.

\section{Threat Model and Methodology}
\label{sec:method}

\subsection{Attacker Model}
\label{subsec:attacker}

Our attacker is a malicious application installed on the same device as the target, holding no permissions beyond those granted automatically at install time (normal-level only).

The attacker installs through normal channels and can send explicit intents to any exported component, query any exported ContentProvider, register intent filters for any custom URI scheme, and bind to any exported Service that carries no \texttt{android:\allowbreak{}permission}. Because Android~14 and later restrict Background Activity Launch (BAL), all Activity evaluations were run from a foreground Activity in the proof-of-concept app, which reflects a realistic attacker capability.

The attacker holds no signature or privileged permissions, cannot modify the system partition or exploit kernel bugs, and cannot reach non-exported components unless granted a temporary URI permission.

Two further ContentProvider observations need capabilities beyond this model and are therefore excluded from the 120 evaluations. First, SettingsProvider skips its per-key read restriction for applications marked test-only, but such an application can be installed only through the Android Debug Bridge (\texttt{adb install -t}). Second, Drive's StorageBackendContentProvider correctly rejects ordinary applications (counted as a hold, CP-18 in Section~\ref{subsec:providers}) but is reachable from the ADB shell, whose user holds the required \texttt{MANAGE\_DOCUMENTS} permission.

\subsection{Evaluation Categories}

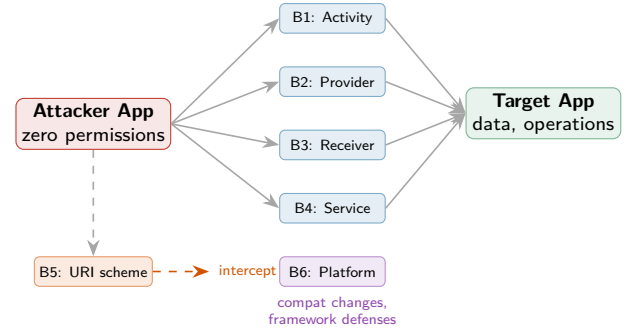
\begin{figure}[t]
\centering
\begin{tikzpicture}[scale=0.93, every node/.style={transform shape},
  node distance=0.6cm,
  appbox/.style={draw, rounded corners=3pt, minimum width=2.2cm, minimum height=0.55cm, font=\scriptsize\sffamily, align=center, inner sep=2pt},
  bndbox/.style={draw, rounded corners=2pt, minimum width=1.5cm, minimum height=0.42cm, font=\tiny\sffamily, align=center, inner sep=2pt, fill=infocolor!12, draw=infocolor!60},
  tgtbox/.style={draw, rounded corners=3pt, minimum width=2.2cm, minimum height=0.55cm, font=\scriptsize\sffamily, align=center, inner sep=2pt},
  arr/.style={-{Stealth[length=2.2mm]}, semithick, gray!70}
]
\node[appbox, fill=gapcolor!12, draw=gapcolor] (atk) at (0,0) {\textbf{Attacker App}\\zero permissions};

\node[bndbox] (b1) at (3.4, 1.5) {B1: Activity};
\node[bndbox] (b2) at (3.4, 0.6) {B2: Provider};
\node[bndbox] (b3) at (3.4,-0.3) {B3: Receiver};
\node[bndbox] (b4) at (3.4,-1.2) {B4: Service};

\node[tgtbox, fill=holdcolor!10, draw=holdcolor!60] (tgt) at (6.4, 0.15) {\textbf{Target App}\\data, operations};

\node[bndbox, fill=warncolor!12, draw=warncolor!60] (b5) at (0,-2.1) {B5: URI scheme};
\node[bndbox, fill=platcolor!12, draw=platcolor!60] (b6) at (3.4,-2.1) {B6: Platform};

\draw[arr] (atk.east) -- (b1.west);
\draw[arr] (atk.east) -- (b2.west);
\draw[arr] (atk.east) -- (b3.west);
\draw[arr] (atk.east) -- (b4.west);
\draw[arr] (b1.east) -- (tgt.west);
\draw[arr] (b2.east) -- (tgt.west);
\draw[arr] (b3.east) -- (tgt.west);
\draw[arr] (b4.east) -- (tgt.west);

\draw[arr, dashed] (atk.south) -- (b5.north);
\draw[arr, dashed, warncolor] (b5.east) -- ++(0.8,0) node[right, font=\tiny\sffamily, text=warncolor] {intercept};

\node[font=\tiny\sffamily, text=platcolor, align=center] at (3.4,-2.65) {compat changes,\\framework defenses};
\end{tikzpicture}
\caption{Threat model. Solid arrows: direct IPC (B1--B4). Dashed: URI scheme interception~(B5). B6 covers platform controls.}
\label{fig:threatmodel}
\end{figure}

We organize evaluations into five inter-component interface categories and one platform security-control category (Figure~\ref{fig:threatmodel}):

\begin{itemize}[leftmargin=1.4em, itemsep=1pt]
\item \textbf{B1} (Activity launch): Does the target Activity verify the caller before processing attacker-controlled extras?
\item \textbf{B2} (ContentProvider query): Does the provider validate query parameters and enforce access control?
\item \textbf{B3} (Broadcast delivery): Does the receiver verify the sender before acting on broadcast data?
\item \textbf{B4} (Service binding): Does the service verify the caller's identity and authorization?
\item \textbf{B5} (URI scheme): Can the platform verify which app should handle a given custom URI scheme?
\item \textbf{B6} (Platform controls): Are platform-level security hardening mechanisms active on the test device?
\end{itemize}

B1--B5 evaluate enforcement at inter-component interfaces. B6 evaluates platform security guardrails that are intended to reinforce component-level enforcement; these are not IPC boundaries themselves but platform-level controls with security intent.

\subsection{Expected-Versus-Observed Framework}

For each evaluation we set the \textbf{expected enforcement} from one or more sources, in decreasing weight: (1)~an explicit platform or security specification; (2)~Android developer documentation describing the intended behavior; (3)~protection-level semantics, when a permission's protection level does not match the sensitivity of what it guards; (4)~sibling comparison, when component siblings in the same manifest carry a security check but the evaluated component does not. A sibling's check does not by itself prove the evaluated component needs one, so sibling comparison only establishes the expectation. Where it is the only source and our evidence shows no effect beyond reaching the component (A-19, A-25, BR-07), we place the evaluation in a separate tier rather than count it as a confirmed gap (Section~\ref{subsec:landscape}).

We determine the \textbf{observed enforcement} by dynamic testing (running a PoC app on a Pixel~6a with Android~17, capturing logcat output), by static analysis (manifest review and code inspection of decompiled APKs or AOSP source), or by both.

Each evaluation is labeled \textbf{GAP} when observed behavior diverges from expected enforcement, or \textbf{HOLDS} when enforcement works as designed. Evidence is tagged S (static only), D (dynamic only), or S+D (both).

Each evaluation carries an identifier whose prefix names its category: A (Activity), CP (ContentProvider), BR (BroadcastReceiver), SV (Service), URI (custom URI scheme), SYS (platform control), and NEG (well-defended component used as a positive control). For example, CP-06 is the sixth ContentProvider evaluation. Section~\ref{sec:results} discusses evaluations in functional groups and cites these identifiers so that each statement can be matched to a specific evaluation, and Appendix~\ref{app:index} lists all 120 evaluations by identifier with the component, package, outcome, and evidence type. Identifiers are numbered consecutively within each category, except that ContentProvider identifiers run to CP-34 although 29 are counted: three duplicate positive controls NEG-06 to NEG-08, and two require ADB access (Section~\ref{subsec:attacker}).

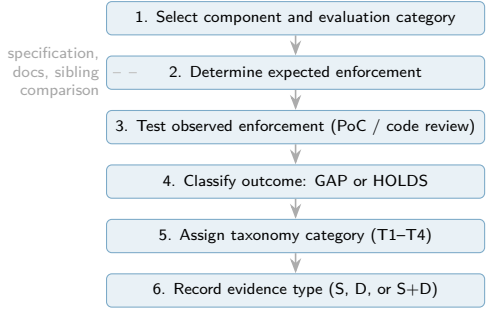
\begin{figure}[t]
\centering
\begin{tikzpicture}[
  wfbox/.style={draw, rounded corners=2pt, fill=infocolor!10, draw=infocolor!60, minimum width=5.0cm, minimum height=0.44cm, font=\tiny\sffamily, align=center, inner sep=2pt},
  arr/.style={-{Stealth[length=2mm]}, semithick, gray!65}]
\node[wfbox] (s1) at (0,0) {1. Select component and evaluation category};
\node[wfbox] (s2) at (0,-0.72) {2. Determine expected enforcement};
\node[wfbox] (s3) at (0,-1.44) {3. Test observed enforcement (PoC / code review)};
\node[wfbox] (s4) at (0,-2.16) {4. Classify outcome: GAP or HOLDS};
\node[wfbox] (s5) at (0,-2.88) {5. Assign taxonomy category (T1--T4)};
\node[wfbox] (s6) at (0,-3.60) {6. Record evidence type (S, D, S+D) and reporting tier};
\foreach \a/\b in {s1/s2,s2/s3,s3/s4,s4/s5,s5/s6}
  \draw[arr] (\a.south) -- (\b.north);
\node[font=\tiny\sffamily, text=gray!70, align=right] at (-3.2,-0.72) {specification,\\docs, sibling\\comparison};
\draw[thin, gray!50, dashed] (-2.1,-0.72) -- (s2.west);
\end{tikzpicture}
\caption{Per-evaluation workflow. Steps 2--6 are applied to each of the 120 evaluations. Expected enforcement is derived from specification, documentation, protection-level semantics, or sibling comparison.}
\label{fig:workflow}
\end{figure}

Figure~\ref{fig:workflow} summarizes the study workflow.

\subsection{Evaluation Scope}
\label{subsec:scope}

\textbf{In scope:} Inter-component interfaces on a Google Pixel~6a running Android~17 (API~37, builds CP2A.260605.012 and CP31.260608.007). Google and AOSP first-party applications, including Chrome and Fitbit.

\textbf{Out of scope:} Wear~OS components, kernel or Binder-layer analysis, network-level attacks, and WebView/JavaScript bridge security except for SYS-10 (Section~\ref{subsec:platform}).

\subsection{Limitations of Methodology}
\label{subsec:methodlimits}

The evaluations are not a random sample. Boundaries were chosen based on where the researcher expected to find gaps, so the gap share describes that selection rather than how often Android IPC boundaries fail in general. We therefore do not report a gap rate. Exactly 21 evaluations (17.5\%) rest on static analysis only; they have no runtime confirmation. All evaluations use a single device (Pixel~6a) running Android~17. The scope is Google and AOSP first-party apps; results may not carry over to third-party applications. App versions are whatever was installed at evaluation time, and Google apps are updated frequently.

\section{Results}
\label{sec:results}

Throughout this section, identifiers in parentheses, such as (CP-06) or (A-12), point to individual evaluations; Appendix~\ref{app:index} gives the full entry for each.

\subsection{Enforcement Landscape}
\label{subsec:landscape}

Table~\ref{tab:outcomes} shows outcomes by category. Of 120 evaluations, 97 record a divergence and 23 record enforcement that held. The counts reflect what was selected, and some categories are uniform by construction. Every custom URI scheme counts as a divergence because the platform cannot verify any custom scheme (Section~\ref{subsec:uri}), so the 18 registrations repeat a single architectural property 18 times. The 10 platform evaluations all record compatibility changes that were either disabled or gated, since enabled changes were not recorded as evaluations; the 100\% figure for B6 describes what was recorded, not an enforcement rate. Among ContentProviders and Services, where both outcomes appeared, holds make up 27.6\% and 28.6\% of evaluations respectively.

\paragraph{Reporting tiers.} Not every divergence is an exploitable flaw, so each is placed in one of six tiers (Table~\ref{tab:tiers}; Appendix~\ref{app:index} lists every evaluation's tier). A \emph{dynamically confirmed gap} is one where device testing showed the component acting on a request from our zero-permission app that the expected enforcement should have blocked; severity varies, and confirmation is of the gap, not of a complete exploit chain. \emph{Missing check, impact not shown} applies when the expectation rests only on sibling comparison and our evidence shows no effect beyond reaching the component. \emph{Static only} means the evidence is code review without runtime confirmation. \emph{Availability only} means the observed effect is a process crash; such issues are typically rated low severity. \emph{Hardening or by-design} covers behavior that is documented or deliberate platform policy: the 18 custom URI schemes, the seven compatibility changes (plus CP-04, which records the same disabled change as SYS-01), and A-07, where \texttt{android:\allowbreak{}autoVerify} behaves exactly as documented. SYS-09 is another researcher's CVE, included for context only.

\begin{table}[!tbp]
\centering \footnotesize
\caption{Reporting tiers for the 97 divergences.}
\label{tab:tiers}
\begin{tabular}{@{}>{\raggedright\arraybackslash}p{2.6cm}r>{\raggedright\arraybackslash}p{3.7cm}@{}}
\toprule
\textbf{Tier} & \textbf{N} & \textbf{Evaluations} \\
\midrule
Dynamically confirmed gap & 54 & 19 CP, 16 A, 10 BR, 8 SV, SYS-08 \\
Missing check, impact not shown & 3 & A-19, A-25, BR-07 \\
Static only & 7 & A-16, A-17, A-27, CP-10, SV-02, SV-10, SYS-10 \\
Availability only (crash) & 5 & A-18, A-20--A-23 \\
Hardening or by-design & 27 & URI-01--URI-18, SYS-01--SYS-07, CP-04, A-07 \\
External CVE (context) & 1 & SYS-09 \\
\midrule
\textbf{Total} & \textbf{97} & \\
\bottomrule
\end{tabular}
\end{table}

\paragraph{Stricter counting.} Table~\ref{tab:strict} shows how totals shift under three progressively stricter rules: removing the external CVE, counting CP-04 and SYS-01 once, and treating the 18 URI registrations as one architectural property. Section~\ref{sec:taxonomy} uses the recorded counts and notes where the choice matters.

\begin{table}[!tbp]
\centering \footnotesize
\caption{Totals under successively stricter counting.}
\label{tab:strict}
\begin{tabular}{@{}>{\raggedright\arraybackslash}p{4.8cm}rr@{}}
\toprule
\textbf{Counting rule} & \textbf{Evals} & \textbf{Gaps} \\
\midrule
As recorded & 120 & 97 \\
Exclude SYS-09 (another researcher's CVE) & 119 & 96 \\
Also count CP-04 and SYS-01 once & 118 & 95 \\
Also count the 18 URI schemes as one property & 101 & 78 \\
\bottomrule
\end{tabular}
\end{table}

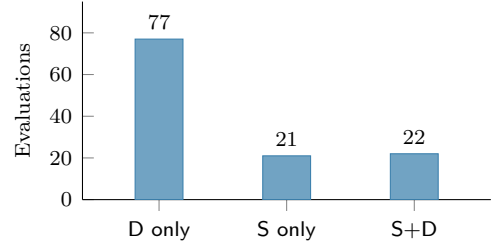
\begin{figure}[t]
\centering
\begin{tikzpicture}
\begin{axis}[
  ybar,
  bar width=18pt,
  width=0.85\columnwidth,
  height=4.2cm,
  ylabel={\scriptsize Evaluations},
  ylabel style={font=\scriptsize},
  symbolic x coords={D only, S only, S+D},
  xtick=data,
  xticklabel style={font=\scriptsize\sffamily},
  yticklabel style={font=\scriptsize},
  ymin=0, ymax=95,
  axis x line*=bottom, axis y line*=left,
  nodes near coords,
  nodes near coords style={font=\scriptsize\bfseries},
  enlarge x limits=0.3
]
\addplot[fill=infocolor!65,draw=infocolor!90] coordinates {(D only,77)(S only,21)(S+D,22)};
\end{axis}
\end{tikzpicture}
\caption{Evidence type distribution. D=dynamic PoC only, S=static analysis only, S+D=both. 82.5\% of evaluations have dynamic evidence.}
\label{fig:evidence}
\end{figure}

Figure~\ref{fig:evidence} shows the evidence breakdown. Dynamic evidence (D or S+D) underlies 99 of 120 evaluations (82.5\%); the remaining 21 (17.5\%) rely on static analysis alone.

\begin{table}[!tbp]
\centering \footnotesize
\caption{Outcomes by evaluation category. Gaps are all recorded divergences; Conf.\ is the subset in the dynamically confirmed tier (Table~\ref{tab:tiers}). Counts reflect what was selected for evaluation, not prevalence.}
\label{tab:outcomes}
\begin{tabular}{@{}p{2.6cm}rrrr@{}}
\toprule
\textbf{Category} & \textbf{N} & \textbf{Gaps} & \textbf{Conf.} & \textbf{Holds} \\
\midrule
B1: Activities         & 27 & 27 & 16 & 0  \\
B2: ContentProv.       & 29 & 21 & 19 & 8  \\
B3: Broadcast          & 11 & 11 & 10 & 0  \\
B4: Services           & 14 & 10 & 8  & 4  \\
B5: URI Schemes        & 18 & 18 & 0  & 0  \\
B6: Platform Ctrl.     & 10 & 10 & 1  & 0  \\
Positive Controls      & 11 & 0  & 0  & 11 \\
\midrule
\textbf{Total}         & \textbf{120} & \textbf{97} & \textbf{54} & \textbf{23} \\
\bottomrule
\end{tabular}
\end{table}

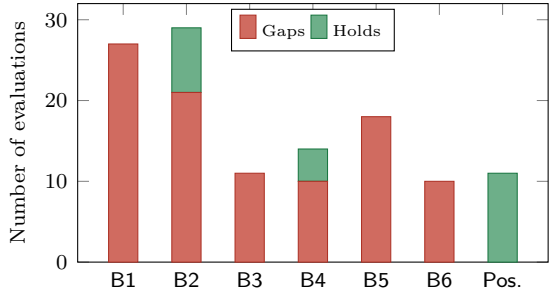
\begin{figure}[t]
\centering
\begin{tikzpicture}
\begin{axis}[
  ybar stacked,
  bar width=11pt,
  width=0.95\columnwidth,
  height=5.0cm,
  ylabel={\scriptsize Number of evaluations},
  ylabel style={font=\scriptsize},
  symbolic x coords={B1,B2,B3,B4,B5,B6,Pos.},
  xtick=data,
  xticklabel style={font=\scriptsize\sffamily},
  yticklabel style={font=\scriptsize},
  ymin=0, ymax=32,
  legend style={at={(0.5,0.98)},anchor=north,font=\tiny,legend columns=2},
  enlarge x limits=0.12
]
\addplot[fill=gapcolor!75,draw=gapcolor!90!black] coordinates {(B1,27)(B2,21)(B3,11)(B4,10)(B5,18)(B6,10)(Pos.,0)};
\addplot[fill=holdcolor!65,draw=holdcolor!90!black] coordinates {(B1,0)(B2,8)(B3,0)(B4,4)(B5,0)(B6,0)(Pos.,11)};
\legend{Gaps, Holds}
\end{axis}
\end{tikzpicture}
\caption{Enforcement outcomes by category. B1=Activities, B2=Content\-Providers, B3=Broadcast\-Receivers, B4=Services, B5=URI~Schemes, B6=Platform Controls, Pos.=Positive Controls.}
\label{fig:gaprate}
\end{figure}

Figure~\ref{fig:gaprate} shows outcomes by evaluation category.

\subsection{Activity Boundary Findings (B1)}
\label{subsec:activities}

All 27 evaluated Activity boundaries show enforcement gaps, across eight functional groups.

\textbf{OAuth/authentication callbacks (A-01--A-03).} Three authentication callback Activities in Gmail and GMS accept arbitrary callers. Cse\-Redirect\-Uri\-Receiver\-Activity (Gmail) processes OAuth callback data from any application. Browser\-Sign\-In\-Response\-Handler\-Activity (GMS) handles Firebase Auth callbacks on both HTTPS and HTTP schemes without origin verification. Authzen\-Deeplink\-Handler\-Activity (GMS) handles sign-in challenge links without verifying the caller.

\textbf{Payment Activities (A-04--A-07).} Four Activities in Google Wallet and GMS handle payment-related operations without adequate caller verification. Finish\-Android\-App\-Redirect\-Proxy\-Activity forwards caller-controlled URLs through GMS's process context. Add\-New\-Card\-Through\-Browser\-Activity accepts card-tokenization requests from any caller. Presentation\-Mode\-Activity activates NFC tap-to-pay without lock screen authentication. Wallet's DeepLinkActivity declares \texttt{android:\allowbreak{}autoVerify}, which by documented design governs only implicit intents, so any application can open its deep-link routes with an explicit intent (A-07). The attribute behaves as specified; we record A-07 as a by-design observation because the exported routes reach sensitive Wallet screens without a caller check.

\textbf{Credential and identity Activities (A-08, A-09).} SharePasswordsActivity (GMS) accepts an arbitrary requesting agent name and package in its credential-sharing flow, and a diagnostics tool's login Activity uses an authentication check that always succeeds.

\textbf{Debug Activities (A-10, A-11).} CalendarDebugActivity runs in the CalendarProvider process, allowing any caller to bypass the {\small\texttt{READ\_CALENDAR}} permission check. ContactsDumpActivity zips and serves the entire contacts data directory without access control.

\textbf{Confused deputy (A-12--A-17).} Six Activities act on a caller's behalf with privileges the caller lacks, the confused deputy pattern~\cite{felt2011}. Three render protected data directly: LaunchInfoActivity renders calendar events without requiring {\small\texttt{READ\_CALENDAR}}, QuickContactActivity renders full contact cards without {\small\texttt{READ\_CONTACTS}}, and HostPhotoPagerActivity renders photos without {\small\texttt{READ\_MEDIA\_IMAGES}}. Three relay caller-controlled requests under the target application's identity: BetterBug's DispatchActivity forwards intents to an Activity protected by the {\small\texttt{DUMP}} permission (A-15), SystemUI's MediaProjectionPermissionActivity reads the package name for its consent-reuse check from an attacker-controlled extra when the caller is unknown (A-16, static evidence), and a Play Store Activity forwards a caller-controlled URI from the Play Store process (A-17, static evidence).

\textbf{DoS and inconsistent validation (A-18--A-23).} Five Activities crash when they receive empty or malformed Intents, causing process-level denial of service (A-18, A-20--A-23). KeyChainActivity (A-20) crashes the system-UID process; BrightnessDialog (A-22) crashes the main SystemUI process. Crash-only issues of this kind are often rated low severity, and we report them in the availability tier rather than as confirmed security gaps. In the same group, a Dialer handler (A-19) is the only one of its sibling handlers that omits the Google signature check the others apply; we did not establish an effect beyond reaching the handler.

\textbf{PII exposure and privileged configuration (A-24--A-27).} An emergency-information screen designed for the lock screen is exported to any application and exposes the owner's name, medical information, and emergency contacts (A-24). Three Activities perform privileged configuration without the permission their context implies: a Wi-Fi calling activation screen whose sibling Activities require {\small\texttt{MODIFY\_PHONE\_STATE}} (A-25), an eSIM download screen that accepts a forced-provisioning flag from any caller (A-26), and a parental-controls setup Activity reachable without a manifest permission (A-27, static evidence).

\subsection{ContentProvider Boundary Findings (B2)}
\label{subsec:providers}

Of 29 evaluated ContentProvider boundaries, 21 show gaps and 8 show holds; ContentProviders and Services are the only interface categories with recorded holds.

\textbf{SQL injection (CP-01--CP-13).} CalendarProvider2 accepts unsanitized input in its selection, projection, and write-path (update and delete) parameters, permitting cross-table data extraction (CP-01--CP-03), and the platform change that would enforce strict query validation for it is disabled (CP-04; see SYS-01). Selection injection also reaches ContactsProvider2, CallLogProvider, and CarrierIdProvider, the last with full \texttt{UNION} injection and no permission required (CP-05, CP-07, CP-08); sortOrder injection affects four providers (CP-09); and CellBroadcastContentProvider applies none of the strict-mode checks (CP-10, static evidence). Where the defenses are applied, they hold: ContactsProvider2 blocks projection injection via {\small\texttt{setStrict(true)}} (CP-06), CalendarProvider2's instances query enables all three strict-mode checks (CP-11), and MediaProvider and DownloadProvider reject SQL keywords in sortOrder (CP-12, CP-13).

\textbf{Shared library gap (CP-14--CP-19).} Four Google Workspace applications (Drive, Docs, Sheets, Slides) share a {\small\texttt{Legacy\-Storage\-Backend\-Content\-Provider}} with zero permission checks on {\small\texttt{query()}} (CP-14--CP-17). The modern {\small\texttt{Storage\-Backend\-Content\-Provider}} in the same packages correctly requires {\small\texttt{MANAGE\_DOCUMENTS}} (CP-18, HOLDS), and the legacy provider's {\small\texttt{openFile()}} resists path traversal because it looks files up by database identifier rather than building a filesystem path from the request (CP-19, HOLDS).

\textbf{Asymmetric permissions (CP-23, CP-24).} MmsSmsProvider declares {\small\texttt{read\-Permission}} but no {\small\texttt{write\-Permission}}, allowing deletion of SMS/MMS records. GoogleSettingsProvider declares {\small\texttt{write\-Permission}} but no {\small\texttt{read\-Permission}}, exposing 18 partner settings.

\textbf{Other provider gaps (CP-20--CP-22, CP-25, CP-27, CP-34).} A debug provider serves a ZIP archive of the contacts database (CP-20); Settings' SuggestionStateProvider reveals screen-lock and biometric enrollment status (CP-21); Messages' AvatarContentProvider proxies arbitrary \texttt{content://} URIs under its own {\small\texttt{READ\_CONTACTS}} permission (CP-22); ServiceStateProvider skips its permission checks for callers targeting API levels below~31 (CP-25); GServicesProvider is guarded only by a normal-level permission while exposing 1,404 configuration entries (CP-27); and a temporary URI grant from the contact picker permits SQL injection that reaches all internal ContactsProvider2 tables (CP-34).

Of the identifiers CP-28 to CP-33, three duplicate positive controls (CP-28 to CP-30) and one falls outside the attacker model (CP-33, Section~\ref{subsec:attacker}). The other two are counted holds: BluetoothOppProvider applies {\small\texttt{setStrict(true)}} and a path permission (CP-31), and the phone application's SIM contact providers require contacts permissions (CP-32).

\subsection{BroadcastReceiver Boundary Findings (B3)}
\label{subsec:receivers}

All 11 evaluated BroadcastReceiver boundaries show gaps.

\textbf{GmsExternalReceiver (BR-01--BR-03).} GMS uses a single exported receiver for 31 distinct Intent actions with no {\small\texttt{android:\allowbreak{}permission}} and no per-action sender validation (BR-03). Behind it, FrpUpdateIntentOperation writes the Factory Reset Protection (FRP) secret without sender verification (BR-01), and AuthZenEventHandler forces re-registration of six CryptAuth security keys (BR-02).

\textbf{Debug receivers (BR-04--BR-06).} TestingToolsBroadcastReceiver is present in 11 Google applications, gated by a runtime Phenotype flag (Google's server-controlled feature-flag system) rather than stripped at build time (BR-04). CalendarProvider's debug receiver responds to a secret-code broadcast (BR-05), and a test receiver in \texttt{com.google.android.apps.scone} forwards simulated state changes to a privileged service (BR-06).

\textbf{Safety-critical receiver (BR-08).} CrisisNotificationBroadcastReceiver accepts broadcast intents from any application, enabling dismissal of earthquake, tsunami, and active shooter alerts.

\textbf{Other receivers (BR-07, BR-09--BR-11).} SystemUI's media-output dialog action is not declared as a protected broadcast although its sibling action is (BR-07); an OMA-DM receiver without a permission invokes a permission-protected receiver directly as a Java method, bypassing the system's broadcast permission check (BR-09); Wallet receivers let any application cancel Wallet notifications, including fraud alerts (BR-10); and the Messages RCS provisioning receiver relies on a normal-level permission and validates the sender for only one of its seven actions (BR-11).

\subsection{Service Boundary Findings (B4)}
\label{subsec:services}

Of 14 evaluated Service boundaries, 10 show gaps and 4 show holds.

\textbf{MediaBrowserService pattern (SV-01, SV-02).} YouTube's MediaBrowserService only blocks SystemUI, accepting all other callers with zero permissions. Bluetooth's MediaBrowserService accepts all callers with zero validation.

\textbf{Missing or weak service permissions (SV-03--SV-06, SV-09, SV-10).} GMS exports 126 services without {\small\texttt{android:\allowbreak{}permission}} (SV-03), and Photos exports seven backup services without permission while keeping its internal backup services unexported (SV-04). Messages and the Google app guard services with normal-level permissions that every application receives automatically (SV-05, SV-10). FederatedCompute performs no Binder caller checks anywhere in its service (SV-06), and an on-device personalization service lacks the permission its sibling service requires (SV-09).

\textbf{Inconsistent AIDL enforcement (SV-07, SV-08).} KeyChainService exposes approximately 50 Binder methods; 5 lack caller identity checks, allowing enumeration of 120 system certificate authority (CA) certificate aliases. NfcService protects all but 2 of approximately 50 methods; the two unprotected methods accept forged host card emulation (HCE) data.

\textbf{Positive controls (SV-11--SV-14).} FelicaServiceImpl enforces {\small\texttt{getCallingUid()}} (SV-11). ZeroPartyBinder performs UID plus signature verification (SV-12). NfcService enforces NFC permissions on all its other methods (SV-13). Two GMS services throw SecurityException on unprivileged bind attempts (SV-14).

\subsection{Custom URI Scheme Findings (B5)}
\label{subsec:uri}

We found 18 custom URI scheme registrations that handle security-sensitive data (URI-01 to URI-18). Where one application registers several related schemes for the same purpose (such as \texttt{tez://} and \texttt{gpay://} in Google Pay), they count as one evaluation. Table~\ref{tab:uri} lists a selection. All 18 share the same architectural limitation: Android cannot verify which application owns a given custom scheme~\cite{rfc8252}, so any installed application can register a competing handler.

\begin{table}[!tbp]
\centering \footnotesize
\caption{Selected custom URI scheme registrations (18 total). All share the architectural limitation that Android cannot verify ownership of non-HTTPS custom schemes~\cite{rfc8252}.}
\label{tab:uri}
\begin{tabular}{@{}>{\raggedright\arraybackslash}p{2.3cm}>{\raggedright\arraybackslash}p{2.0cm}>{\raggedright\arraybackslash}p{2.8cm}@{}}
\toprule
\textbf{Scheme} & \textbf{Application} & \textbf{Data at Risk} \\
\midrule
otpauth:// & Authenticator & TOTP 2FA secret keys \\
LPA:// & eSIM manager & eSIM activation codes \\
FIDO:// & GMS & FIDO2 session data \\
genericidp:// & 5 apps & OAuth authorization codes \\
comgoogle\allowbreak{}wallet:// & GMS (Wallet) & Card tokens, digital car keys \\
tez://, gpay:// & Google Pay & UPI payment data \\
safetyhub:// & Safety Hub & Medical information parameters \\
\bottomrule
\end{tabular}
\end{table}

Across 15 or more packages, these 18 schemes carry authentication credentials, payment tokens, eSIM activation codes, medical information, and other sensitive data. Multiple applications reached for custom schemes independently, even though RFC~8252 had documented the interception risk years earlier.

\subsection{Platform Security-Control Findings (B6)}
\label{subsec:platform}

All 10 platform-level evaluations record a divergence. As Section~\ref{subsec:landscape} notes, only disabled or gated changes were recorded, so this count is not an enforcement rate.

\textbf{Compatibility changes (SYS-01--SYS-07).} We recorded seven compatibility changes with security intent (Table~\ref{tab:compat}). On the test device (Pixel~6a, Android~17), three are disabled and four are on only for applications whose \texttt{targetSdkVersion} is 37 or higher. Gating on \texttt{targetSdkVersion} is Android's deliberate staged-rollout design (Section~\ref{sec:background}), not a defect, and we put these evaluations in the hardening tier. We did not record every evaluated application's \texttt{targetSdkVersion}; AOSP platform applications often target the current SDK, and for those the four gated changes would apply. What our results do show is that the changes are absent for applications that target an earlier SDK, which covers both third-party attacker apps and any first-party app that has not raised its target. Some changes may have different documented defaults; our observations come from \texttt{am compat get-config} on the test device.

\begin{table}[!tbp]
\centering \footnotesize
\caption{Platform compatibility changes with security intent. ``Observed'' reflects the state on the test device (Pixel~6a, Android~17): Disabled means off for all applications; SDK~37 means enabled only for applications whose \texttt{targetSdkVersion} is 37 or higher.}
\label{tab:compat}
\begin{tabular}{@{}l>{\raggedright\arraybackslash}p{3.6cm}l@{}}
\toprule
\textbf{ID} & \textbf{Change} & \textbf{Observed} \\
\midrule
SYS-01 & {\scriptsize ENFORCE\_STRICT\_\allowbreak{}QUERY\_BUILDER} & Disabled \\
SYS-02 & {\scriptsize ENFORCE\_INTENTS\_\allowbreak{}TO\_MATCH\_\allowbreak{}INTENT\_\allowbreak{}FILTERS} & Disabled \\
SYS-03 & {\scriptsize BLOCK\_NULL\_\allowbreak{}ACTION\_INTENTS} & Disabled \\
SYS-04 & {\scriptsize ENFORCE\_STRICT\_\allowbreak{}SQL\_CHECKS} & SDK~37 \\
SYS-05 & {\scriptsize PARCEL\_HARDENING} & SDK~37 \\
SYS-06 & {\scriptsize RESTRICT\_DATA\_\allowbreak{}URI\_COLUMNS} & SDK~37 \\
SYS-07 & {\scriptsize DETECT\_IMPLICIT\_\allowbreak{}URI\_\allowbreak{}PERMISSION\_\allowbreak{}GRANT} & SDK~37 \\
\bottomrule
\end{tabular}
\end{table}

\textbf{Parcelable mismatch regression (SYS-08).} A Settings Tile Parcelable implementation writes and reads fields in different orders, enabling a self-changing Bundle attack. This class of vulnerability (Parcelable serialization/deserialization mismatch) was previously identified in CVE-2023-45777~\cite{cve202345777}; our observation represents a similar pattern on a current platform build. The platform defense ({\small\texttt{PARCEL\_HARDENING}}) is gated at {\small\texttt{targetSdk=37}} on the test device, so it does not protect applications that target an earlier SDK.

\textbf{Third-party library proxy execution (SYS-09).} InCallController in {\small\texttt{system\_\allowbreak{}server}} uses {\small\texttt{create\-Package\-Context()}} with flag~3 ({\small\texttt{CONTEXT\_\allowbreak{}INCLUDE\_\allowbreak{}CODE\,|\allowbreak{}\,CONTEXT\_\allowbreak{}IGNORE\_\allowbreak{}SECURITY}}) against an attacker-controlled package during {\small\texttt{get\-InCall\-Service\-Components}} enumeration, triggered by any app holding the auto-granted {\small\texttt{MANAGE\_\allowbreak{}OWN\_\allowbreak{}CALLS}} permission. According to the original report, this loads code from the attacker's Dalvik Executable (DEX) into \texttt{system\_server} (UID~1000), enabling arbitrary code execution without user interaction. This vulnerability was discovered and reported by sithi~\cite{tlpe2026}, assigned CVE-2026-49881~\cite{cve202649881}, and patched in the September 2026 Android Security Bulletin. We did not discover it and do not count it as our finding (it is in the external tier and excluded under stricter counting, Table~\ref{tab:strict}); we include it as context because our static review confirmed that the vulnerable code path is present in the evaluated builds, which predate that patch. Our search for the same pattern elsewhere found only uses that resolve to Google-signed packages in application context (NEG-09, NEG-10).

\textbf{JavaScript bridge (SYS-10).} The one WebView-related evaluation in scope concerns a JavaScript bridge in Google Maps whose \texttt{openUrl()} method turns script-supplied URLs into native intent launches within the Maps process. This evaluation rests on static evidence only; we did not confirm that an attacker can inject script into the relevant WebView.

\subsection{Positive Controls}
\label{subsec:positivecontrols}

Twenty-three evaluations show enforcement holds. These are calibration cases that confirm the procedure can detect functioning enforcement, not just gaps.

\textbf{Well-defended components (NEG-01--NEG-11).} Eleven evaluations target components chosen as positive controls. Pixel Launcher pairs signature permissions with SHA-256 package allowlists (NEG-01). The NFC stack uses flag~0 for \texttt{createPackageContext}, avoids \texttt{FLAG\_MUTABLE} for PendingIntents, and checks permissions on all Binder methods (NEG-03). ManagedProvisioning verifies callers in code at multiple points even though it is exported without a manifest permission (NEG-04). The strongest components layer manifest permissions, code-level caller checks, and explicit allowlists.

\textbf{ContentProvider and Service defenses.} The eight ContentProvider holds and four Service holds are described in Sections~\ref{subsec:providers} and~\ref{subsec:services}. Table~\ref{tab:posctrl} lists all 23 holds with the mechanism that held in each case.

\begin{table}[!tbp]
\centering \scriptsize
\caption{The 23 enforcement holds and the mechanism that held in each case.}
\label{tab:posctrl}
\begin{tabular}{@{}l>{\raggedright\arraybackslash}p{2.4cm}>{\raggedright\arraybackslash}p{3.65cm}@{}}
\toprule
\textbf{ID} & \textbf{Component} & \textbf{Mechanism that held} \\
\midrule
CP-06 & Contacts\-Provider2 & \texttt{setStrict(true)} blocks projection injection \\
CP-11 & Calendar\-Provider2 (instances) & All three strict-mode checks enabled \\
CP-12 & MediaProvider & Rejects SQL keywords in sortOrder \\
CP-13 & DownloadProvider & Rejects SQL keywords in sortOrder \\
CP-18 & Drive StorageBackend provider & \texttt{MANAGE\_DOCUMENTS} (signature|privileged) \\
CP-19 & Drive legacy \texttt{openFile()} & Lookup by database ID; no path built from input \\
CP-31 & Bluetooth\-Opp\-Provider & \texttt{setStrict(true)} and path permission \\
CP-32 & SIM contact providers & Contacts permissions required \\
\midrule
SV-11 & FelicaServiceImpl & \texttt{getCallingUid()} check \\
SV-12 & ZeroPartyBinder & UID and signature check \\
SV-13 & NfcService (other methods) & NFC permission checks enforced \\
SV-14 & Two GMS services & SecurityException on unprivileged bind \\
\midrule
NEG-01 & Pixel Launcher & Signature permissions and SHA-256 allowlists \\
NEG-02 & Google Meet & No intent forwarding or JavaScript bridge \\
NEG-03 & NFC stack & Safe context flags, immutable PendingIntents, Binder permission checks \\
NEG-04 & Managed\-Provisioning & Code-level caller verification \\
NEG-05 & Phone application & Provider permissions; diagnostics limited to system user and debuggable builds \\
NEG-06 & Gmail providers (10) & SecurityException, empty result, or not exported \\
NEG-07 & GMS providers & Not exported or permission required \\
NEG-08 & FileProviders & Not exported \\
NEG-09 & \texttt{create\-Package\-Context} uses & Resolve only to Google-signed packages \\
NEG-10 & DynamiteModule loaders & Hard-coded GMS package name \\
NEG-11 & 17 other candidates (16 apps) & Properly secured or not exploitable \\
\bottomrule
\end{tabular}
\end{table}

\section{Enforcement Gap Taxonomy}
\label{sec:taxonomy}

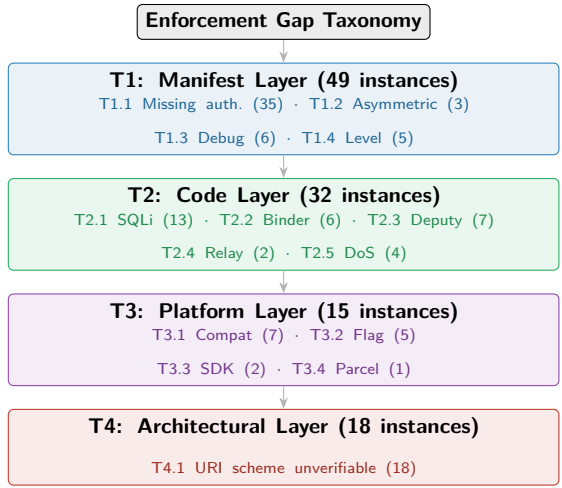
\begin{figure}[t]
\centering
\begin{tikzpicture}[
  root/.style={draw, rounded corners=2pt, fill=gray!15, font=\scriptsize\sffamily\bfseries, inner sep=3pt},
  layerbox/.style={draw, rounded corners=2pt, text width=7.0cm, align=center, inner xsep=4pt, inner ysep=3pt},
  arr/.style={-{Stealth[length=1.5mm]}, thin, gray!60}
]
\node[root] (root) {Enforcement Gap Taxonomy};
\node[layerbox, fill=manifestcolor!10, draw=manifestcolor, below=0.3cm of root] (l1) {{\scriptsize\sffamily\bfseries T1: Manifest Layer (49 instances)}\\[1.5pt]{\tiny\sffamily\color{manifestcolor!80!black}T1.1 Missing auth.\ (35) $\cdot$ T1.2 Asymmetric (3)\\[-1pt]T1.3 Debug (6) $\cdot$ T1.4 Level (5)}};
\node[layerbox, fill=codecolor!10, draw=codecolor, below=0.3cm of l1] (l2) {{\scriptsize\sffamily\bfseries T2: Code Layer (32 instances)}\\[1.5pt]{\tiny\sffamily\color{codecolor!80!black}T2.1 SQLi (13) $\cdot$ T2.2 Binder (6) $\cdot$ T2.3 Deputy (7)\\[-1pt]T2.4 Relay (2) $\cdot$ T2.5 DoS (4)}};
\node[layerbox, fill=platcolor!10, draw=platcolor, below=0.3cm of l2] (l3) {{\scriptsize\sffamily\bfseries T3: Platform Layer (15 instances)}\\[1.5pt]{\tiny\sffamily\color{platcolor!80!black}T3.1 Compat (7) $\cdot$ T3.2 Flag (5)\\[-1pt]T3.3 SDK (2) $\cdot$ T3.4 Parcel (1)}};
\node[layerbox, fill=archcolor!10, draw=archcolor, below=0.3cm of l3] (l4) {{\scriptsize\sffamily\bfseries T4: Architectural Layer (18 instances)}\\[1.5pt]{\tiny\sffamily\color{archcolor!80!black}T4.1 URI scheme unverifiable (18)}};
\draw[arr] (root) -- (l1);
\draw[arr] (l1) -- (l2);
\draw[arr] (l2) -- (l3);
\draw[arr] (l3) -- (l4);
\end{tikzpicture}
\caption{Taxonomy hierarchy: 14 categories across four layers. Instance counts in parentheses; total exceeds 97 due to co-occurring patterns.}
\label{fig:taxonomy}
\end{figure}

The 97 enforcement gaps yield a taxonomy of 14 patterns across four layers. The taxonomy groups failures by where in the enforcement stack they occur. Individual categories cover different kinds of entities (implementation errors, deployment states, architectural limitations), all sharing the same observable effect: an enforcement divergence. Table~\ref{tab:taxonomy} summarizes the taxonomy; Figure~\ref{fig:taxonomy} shows the structure.

\begin{table}[!tbp]
\centering \footnotesize
\caption{14-category enforcement gap taxonomy derived from 97 gaps. Total exceeds 97 because some evaluations exhibit multiple patterns.}
\label{tab:taxonomy}
\begin{tabular}{@{}clcc@{}}
\toprule
\textbf{ID} & \textbf{Category} & \textbf{Count} & \textbf{Layer} \\
\midrule
T1.1 & Missing authorization & 35 & Manifest \\
T1.2 & Asymmetric permission & 3 & Manifest \\
T1.3 & Debug code in production & 6 & Manifest \\
T1.4 & Protection level inadequacy & 5 & Manifest \\
\midrule
T2.1 & SQL injection & 13 & Code \\
T2.2 & Missing Binder check & 6 & Code \\
T2.3 & Confused deputy & 7 & Code \\
T2.4 & Permission trampoline & 2 & Code \\
T2.5 & DoS via input validation & 4 & Code \\
\midrule
T3.1 & Compat change unenforced & 7 & Platform \\
T3.2 & Feature-flag-gated security & 5 & Platform \\
T3.3 & targetSdk gating bypass & 2 & Platform \\
T3.4 & Parcelable mismatch & 1 & Platform \\
\midrule
T4.1 & URI scheme unverifiable & 18 & Arch. \\
\bottomrule
\end{tabular}
\end{table}

T1.1 (``Missing authorization'') covers exported components that handle security-sensitive operations, such as accessing protected data, changing system state, or forwarding privileged actions, without an adequate authorization check. A component that is intentionally public and handles nothing sensitive is not classified here. The remaining categories are as follows.

\emph{T1.2}: a ContentProvider declares a read or write permission but not both. \emph{T1.3}: debug or test components remain exported in release builds. \emph{T1.4}: the guarding permission is normal-level, so every application receives it without prompting.

At the code layer, \emph{T2.1} captures caller-supplied selection, projection, or sort-order text that reaches SQL without validation. \emph{T2.2} covers a Binder method that omits caller identity or permission checks. \emph{T2.3} describes a privileged component accessing a resource for a caller without verifying the caller's own access rights; \emph{T2.4}, an unprotected component forwarding a request to a protected one under its own identity; \emph{T2.5}, missing null or type checks on Intent data that crash the receiving process.

Platform-layer categories: \emph{T3.1}, a compatibility change with security intent is disabled or gated at an SDK level so it misses applications targeting an earlier level. \emph{T3.2}, a security check depends on a server-controlled feature flag. \emph{T3.3}, a permission check applies only to callers targeting a recent SDK level. \emph{T3.4}, a Parcelable writes and reads fields in different orders, allowing a Bundle to change between parses. Finally, \emph{T4.1} describes the architectural case where the platform cannot verify which application owns a custom URI scheme.

Missing authorization on an exported component (T1.1, 35 instances) and unvalidated query parameters reaching SQL (T2.1, 13) are the most common implementation-level patterns. T4.1 has the second largest raw count (18), but records one architectural property seen across 18 registrations; if counted once it would be among the smallest categories. Category counts use the recorded evaluations (Table~\ref{tab:strict} shows what stricter counting does) and span every reporting tier.

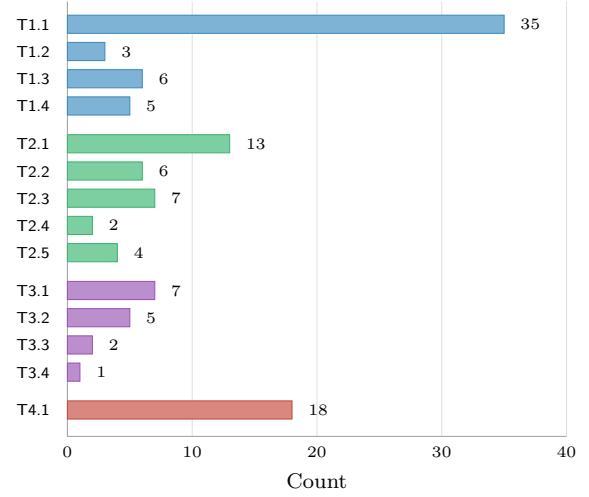
\begin{figure}[t]
\centering
\begin{tikzpicture}[x=0.165cm, y=1cm]
\foreach \v in {0,10,20,30,40} {
  \draw[gray!25, very thin] (\v,0.3) -- (\v,-5.45);
  \node[font=\tiny, below] at (\v,-5.45) {\v};
}
\draw[gray!70] (0,0.3) -- (0,-5.45) -- (40,-5.45);
\node[font=\scriptsize, below] at (20,-5.8) {Count};
\foreach \id/\val/\col/\yy in {
  T1.1/35/manifestcolor/0,    T1.2/3/manifestcolor/-0.36, T1.3/6/manifestcolor/-0.72, T1.4/5/manifestcolor/-1.08,
  T2.1/13/codecolor/-1.58,    T2.2/6/codecolor/-1.94,     T2.3/7/codecolor/-2.30,     T2.4/2/codecolor/-2.66, T2.5/4/codecolor/-3.02,
  T3.1/7/platcolor/-3.52,     T3.2/5/platcolor/-3.88,     T3.3/2/platcolor/-4.24,     T3.4/1/platcolor/-4.60,
  T4.1/18/archcolor/-5.10} {
  \fill[\col!60, draw=\col!90] (0,\yy-0.12) rectangle (\val,\yy+0.12);
  \node[font=\tiny\sffamily, anchor=east] at (-0.6,\yy) {\id};
  \node[font=\tiny, anchor=west] at (\val+0.5,\yy) {\val};
}
\end{tikzpicture}
\caption{Taxonomy category distribution. Colors indicate layer: blue=manifest, green=code, purple=platform, red=architectural.}
\label{fig:taxdist}
\end{figure}

Figure~\ref{fig:taxdist} shows the distribution of taxonomy categories.

\paragraph{Pattern co-occurrence.} Some evaluations show two patterns at once. CalendarProvider2 (CP-01--CP-04) has both SQL injection (T2.1) and a disabled platform defense (T3.1). The Google Workspace legacy providers (CP-14--CP-17) combine SQL injection (T2.1) with feature-flag-gated fixes (T3.2). The Play Store redirect Activity (A-17) and TestingToolsBroadcastReceiver (BR-04) each add a flag-gated fix (T3.2) to their primary pattern. ServiceStateProvider (CP-25) pairs a missing permission with SDK-conditional checks (T3.3). The Settings Tile Parcelable regression (SYS-08) shows Parcelable mismatch (T3.4) alongside the defending compat change \texttt{PARCEL\_\allowbreak{}HARDENING}, which is unenforced (T3.1).

\section{Discussion}
\label{sec:discussion}

\subsection{Summary of Findings by Research Question}

\noindent\textbf{RQ1} (enforcement variation): Outcomes differ across interface categories in our corpus. Holds appeared in ContentProviders (8 of 29) and Services (4 of 14), and in none of the evaluated Activities and BroadcastReceivers. Evaluations were chosen where gaps were expected, so these proportions describe the corpus, not how often Android IPC enforcement fails in general (Section~\ref{subsec:fwdefaults}).

\noindent\textbf{RQ2} (recurring patterns): The taxonomy finds missing authorization on exported components (T1.1, 35 instances) and unvalidated ContentProvider query parameters (T2.1, 13) as the most common implementation-level patterns. The 18 custom URI registrations (T4.1) reflect one architectural property recurring across packages (Section~\ref{sec:taxonomy}).

\noindent\textbf{RQ3} (framework defenses): Categories where the framework offers per-request enforcement primitives show higher holds rates: ContentProviders (\texttt{setStrict} family, separate \texttt{readPermission}/\texttt{writePermission}) at 27.6\% and Services (per-call Binder caller identity) at 28.6\%, against 0\% for evaluated Activities and BroadcastReceivers (Section~\ref{subsec:fwdefaults}).

\subsection{Framework-Provided Defenses and Enforcement Outcomes}
\label{subsec:fwdefaults}

Holds in our corpus appear only in ContentProviders and Services. Both categories give developers per-request enforcement primitives that Activities and BroadcastReceivers lack.

For ContentProviders, those primitives are the three strict-mode methods of \texttt{SQLiteQueryBuilder} (Section~\ref{subsec:cpdefenses}) and separate manifest attributes for read and write permissions. Where developers apply them, enforcement holds: ContactsProvider2, CalendarProvider2's instances query, MediaProvider, DownloadProvider, and BluetoothOppProvider all demonstrate this (CP-06, CP-11--CP-13, CP-31). Services work differently. Every Binder call carries the caller's identity, so a service can check it with \texttt{getCallingUid()} or \texttt{enforceCallingPermission()} on any method it chooses. Three of the four Service holds rely on exactly that (SV-11--SV-13).

Activities are harder to defend. \texttt{getCallingPackage()} works only when the Activity was started for a result; the Android~14 additions \texttt{getLaunchedFromUid()} and \texttt{getLaunchedFromPackage()} report the launcher only when it has opted in to identity sharing via \texttt{ActivityOptions.setShareIdentityEnabled} or already shares the Activity's UID, and otherwise return nothing~\cite{android14compat}. An exported Activity therefore cannot rely on the platform to identify an uncooperative caller, and there is no framework validator for Intent extras comparable to \texttt{setStrict()}. When the launched-from check is applied, it works: SystemUI's MediaProjectionPermissionActivity uses it as its primary check, and the weakness we record (A-16) is confined to the fallback path.

These results are consistent with an association between framework-provided defenses and enforcement outcomes, though the comparison is not controlled. A supplementary triage of all 376 Google and AOSP packages on the test device (3,721 exported components) found no exported ContentProviders with missing access controls beyond those already in the corpus; each was covered by signature-certificate allowlists, calling-package verification, or runtime permission enforcement, which aligns with ContentProvider enforcement benefiting from available framework primitives. Interface categories differ on many dimensions beyond framework support, including API semantics, developer population, and selection criteria, so the data cannot establish a causal relationship.

\subsection{Staged Rollout and the Attacker's Target SDK}
\label{subsec:compatparadox}

On the test device, three of the seven recorded security-related compatibility changes are disabled outright and four activate only at \texttt{targetSdkVersion} 37. Gating on \texttt{targetSdkVersion} is intentional: it keeps existing applications working until their developers raise the target, and we do not argue the policy is wrong. The issue is how it interacts with the attacker model. A hardening change protects an application only after that application has raised its target, and when a change constrains callers rather than providers, enforcement can depend on a target that the attacker chose. As for the three disabled changes, the platform source we reviewed does not record the rationale; they appear to reflect a trade-off made globally on this build.

The same mechanism operates inside applications. ServiceStateProvider applies its permission checks only to callers targeting API~31 or later (CP-25), so an attacker app sidesteps them by declaring an older target such as API~28. The layered effect is visible in CalendarProvider2: it accepts SQL injection (CP-01), and the platform change that would block it, \texttt{ENFORCE\_\allowbreak{}STRICT\_\allowbreak{}QUERY\_\allowbreak{}BUILDER}, is disabled on the test device (SYS-01).

\subsection{Custom URI Schemes: Systematic Reuse of a Known-Vulnerable Pattern}
\label{subsec:uridisc}

RFC~8252 \S8.1, published in October 2017, documents that custom URI scheme redirect URIs can be intercepted and recommends PKCE for public native clients. Nine years on, 18 custom URI scheme registrations carrying sensitive data appear across 15 or more packages. That a well-documented architectural risk keeps being reused suggests the problem is not awareness but incentives: HTTPS App Links, the safer alternative, require more infrastructure to set up. We classify these as architectural exposures rather than necessarily exploitable bugs, since individual applications may apply compensating controls such as PKCE or state validation.

\subsection{Shared Code as Enforcement Gap Amplifier}

A single shared library distributes \texttt{LegacyStorageBackendContentProvider} across four Google Workspace applications, each carrying the same missing permission check (CP-14--CP-17), and TestingToolsBroadcastReceiver appears in 11 applications (BR-04). One gap in a shared component is not one problem; it is that problem multiplied across every application that ships the component. The fix works the same way: patching the library remediates all affected applications at once.

\subsection{The Catch-All Pattern}

GmsExternalReceiver routes 31 Intent actions through one exported receiver with no \texttt{android:\allowbreak{}permission}, relying on each handler to validate its own caller. At least two do not. The approach works only when every handler in that catch-all correctly rejects unauthorized callers, which is a guarantee that grows harder to maintain as the number of actions grows. Pixel Launcher takes the opposite route: per-provider signature permissions with SHA-256 allowlists (NEG-01), so the authorization boundary is at the entry point rather than scattered across handlers.

\subsection{What Distinguishes Well-Defended Components}

Across the 23 holds (Table~\ref{tab:posctrl}), enforcement consistently rests on four mechanisms: manifest permissions at signature or signature|privileged level; code-level caller identity checks; explicit package or signature allowlists rather than general permissions; and framework input validation. The strongest components, Pixel Launcher and the NFC stack, layer several of these together, placing code-level checks on top of manifest permissions. Others were evaluated against a single mechanism, such as the sortOrder validation in MediaProvider and DownloadProvider.

\subsection{Threats to Validity}

\textbf{Selection bias.} Evaluations are not random, and two categories are uniform by construction (custom URI schemes and the recorded platform controls), so we report no overall gap rate. Counts describe the evaluated corpus, not Android IPC in general. A broader measurement study with random sampling would give more informative rates. \textbf{Classification disagreement.} Some divergences may be judged intentional by platform engineers; we put them in the hardening tier, but the line between a hardening observation and a gap is a judgment call. \textbf{Single device and version.} All tests ran on a Pixel~6a with Android~17; compatibility change states may differ on other devices or builds. \textbf{Static-only evidence.} 21 evaluations (17.5\%) have no runtime confirmation. \textbf{Google/AOSP focus.} Third-party application enforcement may look different. \textbf{Temporal.} Specific gaps may be fixed after the evaluation window. \textbf{Confounding.} The association between framework defenses and enforcement outcomes (RQ3) is not a controlled comparison; categories differ on many dimensions beyond framework support.

\section{Related Work}
\label{sec:related}

\textbf{Android IPC security tools.} Three automated tools set the baseline for IPC analysis at scale. ComDroid~\cite{chin2011} applied static analysis to inter-application communication vulnerabilities across 20 applications in detail; CHEX~\cite{lu2012} followed with data-flow analysis for component hijacking over 5{,}486 apps; IccTA~\cite{li2015} extended this with ICC-aware taint tracking. At their scale, our manual study cannot compete. What it adds is an enforcement evaluation with positive controls, cross-category comparison, and platform-level assessment, none of which the automated tools provide.

\textbf{Access-control inconsistency.} A second line of tools looks for the same resource guarded differently on different paths. Kratos~\cite{shao2016} and AceDroid~\cite{aafer2018} compare checks across paths to framework resources; IAceFinder~\cite{zhou2022iace} extends this to Java and native contexts in Android ROMs; Ariadne~\cite{vyas2025} tracks inconsistencies across AOSP and vendor ROMs; FReD~\cite{gorski2022} finds system services that re-delegate file access, a confused deputy at the file layer. All work automatically on the framework, where we work manually on exported application components, using a zero-permission attacker and recording holds alongside gaps. The two approaches meet in our sibling-comparison evaluations, which are a manual form of the inconsistency detection these tools automate.

\textbf{Permission model studies.} Felt et al.~\cite{felt2012} studied how users notice, understand, and respond to Android permission prompts. Bagheri et al.~\cite{bagheri2015} built COVERT, a compositional framework for detecting inter-app permission leakage. Our study tests enforcement of the existing permission model at individual boundaries, not the model's design or user comprehension.

\textbf{ContentProvider security.} Zhou and Jiang~\cite{zhou2013} measured passive content leaks and pollution through exported ContentProviders in 2.3\% of 62,519 applications, establishing the prevalence of the problem. The present study looks at the full enforcement chain instead: app-level injection defenses, framework-level \texttt{setStrict} methods, and platform compatibility changes.

\textbf{Confused deputy.} Felt et al.~\cite{felt2011} named the permission re-delegation attack and proposed IPC Inspection to contain it. Grace et al.~\cite{grace2012} showed the pattern leaking privilege through exported components in preloaded applications on stock Android devices; Bugiel et al.~\cite{bugiel2012} proposed XManDroid for cross-app IPC monitoring. Our contribution here is concrete evaluations in 2026 first-party applications, confirming the pattern is still present in current platform releases.

\textbf{URI scheme security.} RFC~8252 \S8.1~\cite{rfc8252} documents the custom URI scheme interception risk. Liu et al.~\cite{liu2017deeplink} measured deep-link hijacking across more than 160,000 Google Play applications and found that few App Link users passed verification. Our URI evaluations are much narrower: 18 first-party registrations chosen because they carry sensitive data, with no measurement of interception at scale. Their point is that the pattern is still present in first-party code nine years after RFC~8252, not to survey how common it is.

Table~\ref{tab:litcompare} compares our study with prior work across key dimensions.

\begin{table}[!tbp]
\centering \footnotesize
\caption{Comparison with prior work on the dimensions this study focuses on, plus automation and scale, where prior tools are far stronger. \checkmark\,=\,yes, --\,=\,no, $\sim$\,=\,partial.}
\label{tab:litcompare}
\resizebox{\columnwidth}{!}{\begin{tabular}{@{}lcccccc@{}}
\toprule
\textbf{Study} & \rotatebox{70}{\textbf{Pos.\ ctrl.}} & \rotatebox{70}{\textbf{Exp.\ vs.\ obs.}} & \rotatebox{70}{\textbf{Compat state}} & \rotatebox{70}{\textbf{Cross-type}} & \rotatebox{70}{\textbf{URI schemes}} & \rotatebox{70}{\textbf{Automated, at scale}} \\
\midrule
ComDroid \cite{chin2011} & -- & -- & -- & $\sim$ & -- & \checkmark \\
CHEX \cite{lu2012} & -- & -- & -- & \checkmark & -- & \checkmark \\
IccTA \cite{li2015} & -- & -- & -- & \checkmark & -- & \checkmark \\
AceDroid \cite{aafer2018} & -- & $\sim$ & -- & \checkmark & -- & \checkmark \\
Zhou\,\&\,Jiang \cite{zhou2013} & -- & -- & -- & -- & -- & \checkmark \\
Liu et\,al.\ \cite{liu2017deeplink} & -- & -- & -- & -- & \checkmark & \checkmark \\
\textbf{This study} & \checkmark & \checkmark & $\sim$ & \checkmark & $\sim$ & -- \\
\bottomrule
\end{tabular}}
\end{table}

\section{Conclusion}
\label{sec:conclusion}

This paper reports 120 security-enforcement evaluations across five Android inter-component interface categories and platform security controls. The expected-versus-observed framework treats each evaluation as a boundary assessment, recording both divergences (97) and holds (23). The divergences break down into 54 dynamically confirmed gaps, 3 missing checks without shown impact, 7 static-only cases, 5 crash-only availability issues, 27 hardening or by-design observations, and one external CVE. A 14-category taxonomy places them across manifest, code, platform, and architectural layers; missing authorization on exported components (T1.1) and unvalidated ContentProvider query parameters (T2.1) are the most common implementation-level patterns (RQ2).

Three observations stand out. Outside the dedicated positive controls, holds appeared only in ContentProviders (8 of 29) and Services (4 of 14), the two categories where the framework offers per-request enforcement primitives; no evaluated Activity or BroadcastReceiver showed a hold (RQ1, RQ3). The association is not controlled, but it raises a practical question: would stronger framework-level defaults for Activities and BroadcastReceivers produce better outcomes? Compatibility changes with security intent on the test device are either disabled or gated by \texttt{targetSdkVersion}, a deliberate staged-rollout policy that leaves applications at earlier targets outside the change's protection. And 18 custom URI scheme registrations carry sensitive data even though the platform cannot verify custom-scheme ownership, nine years after RFC~8252 documented the interception risk.

Strong enforcement is achievable. The best-defended components layer manifest permissions, code-level caller checks, and explicit allowlists, and the positive controls confirm this combination works. The problem is not that the tools are missing; it is that they are applied unevenly.

\section*{Responsible Disclosure}

\newcommand{\pending}[1]{\textcolor{red}{\textbf{[PENDING: #1]}}}

All evaluated components are maintained by Google or are part of AOSP. SYS-09 was found and disclosed by another researcher and was patched in the September 2026 Android Security Bulletin; we include it for context only. \pending{For each divergence named in Section~\ref{sec:results}, state whether it was reported to the Google Vulnerability Reward Program or the Android security team, the report identifier, the vendor's assessment (accepted, duplicate, intended behavior, or won't fix), and whether a fix has shipped. Where an issue is not yet fixed and the vendor has not agreed to publication, describe it at the level of its vulnerability class only, without component names or reproduction detail.} The proof-of-concept applications are not released.

\section*{Ethical Considerations}

All evaluations were done on the researcher's own device (Google Pixel~6a). No user data was accessed or exfiltrated. Dynamic evaluations used proof-of-concept applications running on the researcher's own accounts. The study examines publicly available system components; every observation comes from manifest inspection, publicly available AOSP source, and standard IPC interactions on a researcher-controlled device.

\section*{Declaration of Generative AI Use}
During the preparation of this work, the author used Claude (Anthropic), a large language model, to draft and edit the manuscript text and to revise its wording for readability, to revise the text in response to reviewer feedback, to check that the numbers reported in the text agree with the released data, and to prepare the LaTeX and Word versions of the manuscript. The study design, the experiments, the findings, and the data are the author's own. The author reviewed and edited all text produced with the tool and takes full responsibility for the content of this publication. The tool is not an author of this work.


\onecolumn
\appendix
\section{Index of All 120 Evaluations}
\label{app:index}
This appendix lists every evaluation counted in the study, grouped by category. The identifier prefix names the category: A (Activity), CP (ContentProvider), BR (BroadcastReceiver), SV (Service), URI (custom URI scheme), SYS (platform control), and NEG (well-defended component used as a positive control). Each row gives the component and the boundary that was tested (for URI schemes, the scheme and the data it carries), the package or scope, the outcome (GAP: observed behavior diverges from expected enforcement; HOLDS: enforcement works as designed), the evidence type (S: static analysis only; D: dynamic proof-of-concept testing on the device; S+D: both), and the reporting tier of Section~\ref{subsec:landscape} (C: dynamically confirmed gap; I: missing check, impact not shown; S: static only; A: availability only; H: hardening or by-design observation; X: external CVE included as context; P: hold). ContentProvider identifiers are not contiguous: CP-26 and CP-33 require ADB access and are excluded (Section~\ref{subsec:attacker}), and CP-28 to CP-30 duplicate positive controls NEG-06 to NEG-08.

{\scriptsize
\begin{longtable}{@{}l>{\raggedright\arraybackslash}p{6.5cm}>{\raggedright\arraybackslash}p{4.4cm}ccc@{}}
\toprule
\textbf{ID} & \textbf{Component and boundary tested} & \textbf{Package or scope} & \textbf{Outcome} & \textbf{Evid.} & \textbf{Tier} \\
\midrule
\endfirsthead
\multicolumn{6}{@{}l}{\textit{Appendix table, continued}} \\
\toprule
\textbf{ID} & \textbf{Component and boundary tested} & \textbf{Package or scope} & \textbf{Outcome} & \textbf{Evid.} & \textbf{Tier} \\
\midrule
\endhead
\bottomrule
\endfoot
\addlinespace[3pt]
\multicolumn{6}{@{}l}{\textbf{B1: Activities (27 evaluations, all gaps)}} \\
\addlinespace[1pt]
A-01 & Cse\-Redirect\-Uri\-Receiver\-Activity: Caller verification on OAuth callback receiver (HTTPS scheme) & com.\allowbreak{}google.\allowbreak{}android.\allowbreak{}gm (Gmail) & GAP & D & C \\
A-02 & Browser\-Sign\-In\-Response\-Handler\-Activity: Caller verification on Firebase Auth callback (HTTPS + HTTP) & com.\allowbreak{}google.\allowbreak{}android.\allowbreak{}gms & GAP & D & C \\
A-03 & Authzen\-Deeplink\-Handler\-Activity: Caller verification on Authzen challenge handler (google.authzen:/\allowbreak{}/\allowbreak{} scheme, noDisplay) & com.\allowbreak{}google.\allowbreak{}android.\allowbreak{}gms & GAP & D & C \\
A-04 & Finish\-Android\-App\-Redirect\-Proxy\-Activity: URL validation on redirect proxy chain (caller-controlled URL laundered through GMS context) & com.\allowbreak{}google.\allowbreak{}android.\allowbreak{}gms (Wallet) & GAP & D & C \\
A-05 & Add\-New\-Card\-Through\-Browser\-Activity: Caller verification on card tokenization handler (comgooglewallet:/\allowbreak{}/\allowbreak{} scheme, singleInstance) & com.\allowbreak{}google.\allowbreak{}android.\allowbreak{}gms (Wallet) & GAP & D & C \\
A-06 & Presentation\-Mode\-Activity: Authentication enforcement for NFC tap-to-pay activation (QUICKDRAW action bypasses lockscreen/\allowbreak{}biometric) & com.\allowbreak{}google.\allowbreak{}android.\allowbreak{}apps.\allowbreak{}walletnfcrel & GAP & D & C \\
A-07 & DeepLinkActivity: Permission enforcement on deep link handler (autoVerify only protects implicit intents) & com.\allowbreak{}google.\allowbreak{}android.\allowbreak{}apps.\allowbreak{}walletnfcrel & GAP & D & H \\
A-08 & Share\-Passwords\-Activity: Caller identity verification on AI agent credential-sharing flow (accepts arbitrary agent\_\allowbreak{}name/\allowbreak{}agent\_\allowbreak{}package extras) & com.\allowbreak{}google.\allowbreak{}android.\allowbreak{}gms & GAP & D & C \\
A-09 & End\-User\-Login\-Activity: Authentication enforcement on priv-app diagnostics (y() returns true unconditionally) & com.\allowbreak{}google.\allowbreak{}android.\allowbreak{}apps.\allowbreak{}diagnosticstool & GAP & S+D & C \\
A-10 & CalendarDebug + Calendar\-Debug\-Activity: Export control on debug activities (bypass READ\_\allowbreak{}CALENDAR by running in provider process) & com.\allowbreak{}android.\allowbreak{}providers.\allowbreak{}calendar & GAP & S+D & C \\
A-11 & Contacts\-Dump\-Activity: Export control on debug activity (zips entire contacts data directory) & com.\allowbreak{}android.\allowbreak{}providers.\allowbreak{}contacts & GAP & D & C \\
A-12 & LaunchInfoActivity: Permission enforcement on exported activity (renders calendar events without READ\_\allowbreak{}CALENDAR; sequential ID enumeration) & com.\allowbreak{}google.\allowbreak{}android.\allowbreak{}calendar & GAP & D & C \\
A-13 & Quick\-Contact\-Activity: Permission enforcement on exported activity (renders full contact cards without READ\_\allowbreak{}CONTACTS) & com.\allowbreak{}google.\allowbreak{}android.\allowbreak{}contacts & GAP & D & C \\
A-14 & Host\-Photo\-Pager\-Activity + EditActivity: Permission enforcement on exported activities (render photos without READ\_\allowbreak{}MEDIA\_\allowbreak{}IMAGES; sequential MediaStore ID enumeration) & com.\allowbreak{}google.\allowbreak{}android.\allowbreak{}apps.\allowbreak{}photos & GAP & D & C \\
A-15 & DispatchActivity: Permission relay/\allowbreak{}trampoline (unpermissioned DispatchActivity forwards intents to DUMP-protected Bug\-Intent\-Dialog\-Activity using BetterBug's identity) & com.\allowbreak{}google.\allowbreak{}android.\allowbreak{}apps.\allowbreak{}betterbug & GAP & D & C \\
A-16 & Media\-Projection\-Permission\-Activity: Caller identity validation (confused deputy: when getCallingPackage()=null, reads package name from attacker-controlled extra for consent-reuse check) & com.\allowbreak{}android.\allowbreak{}systemui & GAP & S & S \\
A-17 & App\-Discovery\-Launch\-Activity: URL validation on intent redirect (singleInstance; forwards caller-controlled URI from Play Store process context; flag-gated fix) & com.\allowbreak{}android.\allowbreak{}vending (Play Store) & GAP & S & S \\
A-18 & ImRootActivity, Deep\-Link\-Entry\-Point\-Activity, Assisted\-Sign\-In\-Activity: Input validation on Chimera module-loading framework (crash loop from empty intents, 5-7+ restarts) & com.\allowbreak{}google.\allowbreak{}android.\allowbreak{}gms & GAP & D & A \\
A-19 & GatewayActivity (Dobby\-Promo\-Gateway\-Handler): Caller verification consistency (Dobby handler is the ONLY GatewayActivity handler lacking Google\-Signature\-Verifier check) & com.\allowbreak{}google.\allowbreak{}android.\allowbreak{}dialer & GAP & S+D & I \\
A-20 & KeyChainActivity: Input validation on Serializable extras (ClassCastException crashes system-UID process) & com.\allowbreak{}android.\allowbreak{}keychain & GAP & D & A \\
A-21 & ReceiverActivity: Input validation on Parcelable extras (NPE crashes android.process.media) & com.\allowbreak{}android.\allowbreak{}mtp & GAP & D & A \\
A-22 & BrightnessDialog: Dead legacy code (unconditionally throws Illegal\-State\-Exception; crashes main SystemUI process) & com.\allowbreak{}android.\allowbreak{}systemui & GAP & D & A \\
A-23 & Key\-Attestation\-Warning\-Activity: Null check on getCallingPackage() (crash loop, 10+ cycles; cascading Play Store billing service restart) & com.\allowbreak{}google.\allowbreak{}android.\allowbreak{}gms & GAP & D & A \\
A-24 & Emergency\-Dialer\-Activity + LockScreenActivity: Caller context validation (designed for lockscreen but exported to any app; exposes full name, medical info, emergency contacts) & com.\allowbreak{}google.\allowbreak{}android.\allowbreak{}apps.\allowbreak{}safetyhub & GAP & D & C \\
A-25 & Wfc\-Activation\-Activity: Permission enforcement (siblings correctly require MODIFY\_\allowbreak{}PHONE\_\allowbreak{}STATE; this activity has none) & com.\allowbreak{}google.\allowbreak{}android.\allowbreak{}wfcactivation & GAP & D & I \\
A-26 & QrDownloadActivity: Caller validation on eSIM provisioning (FORCE\_\allowbreak{}PROVISION=true from zero-perm app) & com.\allowbreak{}google.\allowbreak{}android.\allowbreak{}euicc & GAP & D & C \\
A-27 & KidSetupActivity: Export control on parental controls (accepts DM\_\allowbreak{}PRE\_\allowbreak{}REMOVE\_\allowbreak{}ACCOUNT, ENFORCE\_\allowbreak{}LAUNCHER without manifest permission) & com.\allowbreak{}google.\allowbreak{}android.\allowbreak{}gms.\allowbreak{}supervision & GAP & S & S \\
\addlinespace[3pt]
\multicolumn{6}{@{}l}{\textbf{B2: ContentProviders (29 evaluations: 21 gaps, 8 holds)}} \\
\addlinespace[1pt]
CP-01 & CalendarProvider2 (selection): Input validation on selection parameter (arbitrary subqueries in WHERE clause) & com.\allowbreak{}android.\allowbreak{}providers.\allowbreak{}calendar & GAP & D & C \\
CP-02 & CalendarProvider2 (projection): Input validation on projection parameter (arbitrary SQL functions and subqueries; cross-table extraction in single queries) & com.\allowbreak{}android.\allowbreak{}providers.\allowbreak{}calendar & GAP & D & C \\
CP-03 & CalendarProvider2 (write path): Input validation on UPDATE/\allowbreak{}DELETE selection (side-channel confidentiality breach via write operations) & com.\allowbreak{}android.\allowbreak{}providers.\allowbreak{}calendar & GAP & D & C \\
CP-04 & CalendarProvider2 (compat gate): Platform enforcement of setStrict(true) via ENFORCE\_\allowbreak{}STRICT\_\allowbreak{}QUERY\_\allowbreak{}BUILDER (ChangeId 143231523 globally disabled on Android 17) & com.\allowbreak{}android.\allowbreak{}providers.\allowbreak{}calendar & GAP & S+D & H \\
CP-05 & ContactsProvider2 (selection): Input validation on selection parameter (CASE WHEN subqueries with cross-table references; blind boolean binary search) & com.\allowbreak{}android.\allowbreak{}providers.\allowbreak{}contacts & GAP & D & C \\
CP-06 & ContactsProvider2 (projection): Input validation on projection parameter (correctly blocked by setStrict(true)) & com.\allowbreak{}android.\allowbreak{}providers.\allowbreak{}contacts & HOLDS & D & P \\
CP-07 & CallLogProvider (selection): Input validation on selection parameter (blind boolean; internal tables currently empty) & com.\allowbreak{}android.\allowbreak{}providers.\allowbreak{}contacts & GAP & D & C \\
CP-08 & CarrierIdProvider (selection): Input validation on selection parameter (full UNION injection with 13-column matching; no permission required) & com.\allowbreak{}android.\allowbreak{}providers.\allowbreak{}telephony & GAP & D & C \\
CP-09 & SortOrder injection (multi-provider): Input validation on sortOrder parameter (CASE WHEN subqueries in ORDER BY; 4 providers affected) & Contacts, Calendar, Settings, UserDictionary & GAP & D & C \\
CP-10 & Cell\-Broadcast\-Content\-Provider: Input validation on query parameters (no setStrict; projection/\allowbreak{}selection/\allowbreak{}sortOrder injection possible; gated by READ\_\allowbreak{}CELL\_\allowbreak{}BROADCASTS) & com.\allowbreak{}google.\allowbreak{}android.\allowbreak{}cellbroadcastreceiver & GAP & S & S \\
CP-11 & CalendarProvider2 (instances): Input validation on instances query (setStrict + setStrictColumns + setStrictGrammar all enabled unconditionally) & com.\allowbreak{}android.\allowbreak{}providers.\allowbreak{}calendar & HOLDS & S+D & P \\
CP-12 & MediaProvider: sortOrder validation (properly rejects ``Invalid token SELECT'') & (system) & HOLDS & D & P \\
CP-13 & DownloadProvider: sortOrder validation (properly rejects ``Invalid token SELECT'') & (system) & HOLDS & D & P \\
CP-14 & Legacy\-Storage\-Backend\-Content\-Provider: Permission enforcement on exported provider (query(): zero checks; openFile()/\allowbreak{}call(): gated by Phenotype flag 45793750, defaults false) & com.\allowbreak{}google.\allowbreak{}android.\allowbreak{}apps.\allowbreak{}docs (Drive) & GAP & S+D & C \\
CP-15 & Legacy\-Storage\-Backend\-Content\-Provider: Same shared library vulnerability as CP-14 & com.\allowbreak{}google.\allowbreak{}android.\allowbreak{}apps.\allowbreak{}docs.\allowbreak{}editors.\allowbreak{}docs (Docs) & GAP & S+D & C \\
CP-16 & Legacy\-Storage\-Backend\-Content\-Provider: Same shared library vulnerability as CP-14 & com.\allowbreak{}google.\allowbreak{}android.\allowbreak{}apps.\allowbreak{}docs.\allowbreak{}editors.\allowbreak{}trix (Sheets) & GAP & D & C \\
CP-17 & Legacy\-Storage\-Backend\-Content\-Provider: Same shared library vulnerability as CP-14 & com.\allowbreak{}google.\allowbreak{}android.\allowbreak{}apps.\allowbreak{}docs.\allowbreak{}editors.\allowbreak{}slides (Slides) & GAP & D & C \\
CP-18 & Storage\-Backend\-Content\-Provider: Permission enforcement on SAF provider (MANAGE\_\allowbreak{}DOCUMENTS, signature|privileged) & com.\allowbreak{}google.\allowbreak{}android.\allowbreak{}apps.\allowbreak{}docs (Drive) & HOLDS & S+D & P \\
CP-19 & Legacy\-Storage\-Backend openFile(): Path traversal via URI manipulation (database ID lookup, not filesystem path construction; UriMatcher blocks multi-segment traversal) & (Drive/\allowbreak{}Docs/\allowbreak{}Sheets/\allowbreak{}Slides) & HOLDS & S & P \\
CP-20 & DumpFileProvider: Export control on debug provider (serves contacts database ZIP; no permission) & com.\allowbreak{}android.\allowbreak{}providers.\allowbreak{}contacts & GAP & D & C \\
CP-21 & Suggestion\-State\-Provider: Permission enforcement (no permission, no getCallingPackage() check; exposes screen lock and biometric enrollment status) & com.\allowbreak{}android.\allowbreak{}settings & GAP & D & C \\
CP-22 & Avatar\-Content\-Provider: URI validation on proxy provider (inverted allowlist proxies any content:/\allowbreak{}/\allowbreak{} URI through Messages' READ\_\allowbreak{}CONTACTS) & com.\allowbreak{}google.\allowbreak{}android.\allowbreak{}apps.\allowbreak{}messaging & GAP & D & C \\
CP-23 & MmsSmsProvider: Write permission declaration (readPermission=READ\_\allowbreak{}SMS but no writePermission; zero-perm app can delete all SMS/\allowbreak{}MMS) & com.\allowbreak{}android.\allowbreak{}providers.\allowbreak{}telephony & GAP & D & C \\
CP-24 & Google\-Settings\-Provider: Read permission declaration (writePermission set but no readPermission; leaks 18 partner settings including privacy prefs) & com.\allowbreak{}google.\allowbreak{}settings & GAP & S+D & C \\
CP-25 & Service\-State\-Provider: targetSdkVersion gating on permission checks (targetSdk < 31 bypasses both READ\_\allowbreak{}PRIVILEGED\_\allowbreak{}PHONE\_\allowbreak{}STATE and location permission) & com.\allowbreak{}android.\allowbreak{}phone & GAP & D & C \\
CP-27 & GServicesProvider: Permission protection level adequacy (READ\_\allowbreak{}GSERVICES is normal, auto-granted; 1404 config entries including 3 API keys, 86 URLs, 10 certs) & com.\allowbreak{}google.\allowbreak{}android.\allowbreak{}gsf & GAP & D & C \\
CP-31 & Bluetooth\-Opp\-Provider: SQL injection defense (setStrict(true) + path-permission) & com.\allowbreak{}google.\allowbreak{}android.\allowbreak{}bluetooth & HOLDS & S & P \\
CP-32 & IccProvider, Sim\-Phonebook\-Provider: Permission enforcement (READ/\allowbreak{}WRITE\_\allowbreak{}CONTACTS required) & com.\allowbreak{}android.\allowbreak{}phone & HOLDS & S & P \\
CP-34 & ContactsProvider2 zero-perm variant: URI grant scope (ACTION\_\allowbreak{}PICK contact picker grants temporary URI but SQLi from that URI accesses ALL internal tables) & com.\allowbreak{}android.\allowbreak{}providers.\allowbreak{}contacts & GAP & D & C \\
\addlinespace[3pt]
\multicolumn{6}{@{}l}{\textbf{B3: BroadcastReceivers (11 evaluations, all gaps)}} \\
\addlinespace[1pt]
BR-01 & Gms\-External\-Receiver (FRP): Per-action caller validation (Frp\-Update\-Intent\-Operation writes FRP secret without verification) & com.\allowbreak{}google.\allowbreak{}android.\allowbreak{}gms & GAP & D & C \\
BR-02 & Gms\-External\-Receiver (CryptAuth): Per-action caller validation (Auth\-Zen\-Event\-Handler triggers ForceRegistration of 6 security keys) & com.\allowbreak{}google.\allowbreak{}android.\allowbreak{}gms & GAP & D & C \\
BR-03 & Gms\-External\-Receiver (systemic): Architectural pattern: single receiver, 31 actions, no android:permission, no per-action validation & com.\allowbreak{}google.\allowbreak{}android.\allowbreak{}gms & GAP & D & C \\
BR-04 & Testing\-Tools\-Broadcast\-Receiver: Build-time stripping of debug code (receiver exported in production; gated by runtime Phenotype flag, not build-time removal) & 11 Google apps (Gmail, Dialer, Files, Keep, etc.) & GAP & S+D & C \\
BR-05 & Calendar\-Debug\-Receiver: Export control on debug receiver (accepts SECRET\_\allowbreak{}CODE broadcast) & com.\allowbreak{}android.\allowbreak{}providers.\allowbreak{}calendar & GAP & S+D & C \\
BR-06 & TestReceiver: Export control on test receiver (no permission, no build-type guard; forwards SIMULATE\_\allowbreak{}STATE to privileged StateService) & com.\allowbreak{}google.\allowbreak{}android.\allowbreak{}apps.\allowbreak{}scone & GAP & S+D & C \\
BR-07 & Media\-Output\-Dialog\-Receiver: Protected broadcast declaration (LAUNCH\_\allowbreak{}MEDIA\_\allowbreak{}OUTPUT\_\allowbreak{}DIALOG NOT declared as protected; sibling LAUNCH\_\allowbreak{}SYSTEM\_\allowbreak{}MEDIA\_\allowbreak{}OUTPUT\_\allowbreak{}DIALOG IS; no validation of package\_\allowbreak{}name/\allowbreak{}user\_\allowbreak{}handle extras) & com.\allowbreak{}android.\allowbreak{}systemui & GAP & S+D & I \\
BR-08 & Crisis\-Notification\-Broadcast\-Receiver: Authorization for safety-critical notification dismissal (any app can dismiss earthquake/\allowbreak{}tsunami/\allowbreak{}active shooter alerts) & com.\allowbreak{}google.\allowbreak{}android.\allowbreak{}apps.\allowbreak{}safetyhub & GAP & S+D & C \\
BR-09 & Telephony\-Broadcast\-Receiver: Permission enforcement path (unpermissioned receiver directly calls DMIntentReceiver.onReceive() as Java method, bypassing OS broadcast permission check) & com.\allowbreak{}android.\allowbreak{}omadm.\allowbreak{}service & GAP & D & C \\
BR-10 & Gms\-Core\-Delegate\-Receiver + Notification\-Broadcast\-Receiver: Authorization for notification management (CANCEL\_\allowbreak{}NOTIFICATIONS suppresses fraud alerts; Notification\-Broadcast\-Receiver proxies arbitrary PendingIntents) & com.\allowbreak{}google.\allowbreak{}android.\allowbreak{}apps.\allowbreak{}walletnfcrel & GAP & D & C \\
BR-11 & Rcs\-Provisioning\-Broadcast\-Receiver: Permission protection level (PROVISIONING\_\allowbreak{}EVENT is normal, auto-granted; validates sender only for 1 of 7 actions) & com.\allowbreak{}google.\allowbreak{}android.\allowbreak{}apps.\allowbreak{}messaging & GAP & D & C \\
\addlinespace[3pt]
\multicolumn{6}{@{}l}{\textbf{B4: Services (14 evaluations: 10 gaps, 4 holds)}} \\
\addlinespace[1pt]
SV-01 & Main\-App\-Media\-Browser\-Service: Caller validation in onGetRoot() (only blocks SystemUI; accepts all other callers; zero permissions for metadata + playback control) & com.\allowbreak{}google.\allowbreak{}android.\allowbreak{}youtube & GAP & S+D & C \\
SV-02 & Bluetooth\-Media\-Browser\-Service: Caller validation in onGetRoot() (accepts ALL callers with zero validation; returns BrowseTree.ROOT) & com.\allowbreak{}google.\allowbreak{}android.\allowbreak{}bluetooth & GAP & S & S \\
SV-03 & 126 exported services: Manifest-level permission enforcement (126 services exported with no android:permission; FRP service returns 56-byte config blob) & com.\allowbreak{}google.\allowbreak{}android.\allowbreak{}gms & GAP & S+D & C \\
SV-04 & 7 backup services: Permission enforcement (all exported=true, no permission; internal backup services correctly exported=false) & com.\allowbreak{}google.\allowbreak{}android.\allowbreak{}apps.\allowbreak{}photos & GAP & D & C \\
SV-05 & Grpc\-Endpoint\-Service: Permission protection level (ACCESS\_\allowbreak{}ENDPOINT\_\allowbreak{}SERVICE is normal, auto-granted) & com.\allowbreak{}google.\allowbreak{}android.\allowbreak{}apps.\allowbreak{}messaging & GAP & D & C \\
SV-06 & Federated\-Compute\-Managing\-Service\-Impl: Binder caller identity check (zero enforce\-Calling\-Permission/\allowbreak{}checkCallingOrSelf/\allowbreak{}getCallingUid in entire codebase; schedule()/\allowbreak{}cancel() accept caller-supplied identity) & com.\allowbreak{}google.\allowbreak{}android.\allowbreak{}federatedcompute & GAP & S+D & C \\
SV-07 & KeyChainService (5 of \textasciitilde{}50 methods): Per-method authorization (5 methods including CA enumeration lack caller check; 120 system CA aliases + DER certs readable) & com.\allowbreak{}android.\allowbreak{}keychain & GAP & D & C \\
SV-08 & NfcService (notifyTestHceData, notify\-Hce\-Deactivated): Per-method authorization (2 methods lack NFC permission check; forged HCE APDU injection into card-emulation pipeline) & com.\allowbreak{}google.\allowbreak{}android.\allowbreak{}nfc & GAP & D & C \\
SV-09 & Odp\-Result\-Handling\-Service: Permission enforcement (no permission; sibling correctly requires BIND\_\allowbreak{}EXAMPLE\_\allowbreak{}STORE\_\allowbreak{}SERVICE; writes attacker-controlled tokens under spoofed ComponentName) & com.\allowbreak{}google.\allowbreak{}android.\allowbreak{}ondevicepersonalization.\allowbreak{}services & GAP & D & C \\
SV-10 & LensService: Permission protection level (LENS\_\allowbreak{}SERVICE is normal, auto-granted; telemetry suggests startActivity and PendingIntent creation on behalf of caller) & com.\allowbreak{}google.\allowbreak{}android.\allowbreak{}googlequicksearchbox & GAP & S & S \\
SV-11 & FelicaServiceImpl: Binder caller UID check (getCallingUid() properly enforced) & com.\allowbreak{}google.\allowbreak{}android.\allowbreak{}apps.\allowbreak{}walletnfcrel & HOLDS & S & P \\
SV-12 & ZeroPartyBinder (GMS Supervision): Caller validation (UID check + signature check; only GMS can bind) & com.\allowbreak{}google.\allowbreak{}android.\allowbreak{}gms.\allowbreak{}supervision & HOLDS & S & P \\
SV-13 & NfcService (all other Binder methods): Permission enforcement (NFC, NFC admin, foreground checks all enforced) & com.\allowbreak{}google.\allowbreak{}android.\allowbreak{}nfc & HOLDS & S & P \\
SV-14 & GMS Hardware\-Info\-Query\-Service, GooglePowerService: AIDL-level authorization (SecurityException on unprivileged bind) & com.\allowbreak{}google.\allowbreak{}android.\allowbreak{}gms & HOLDS & D & P \\
\addlinespace[3pt]
\multicolumn{6}{@{}l}{\textbf{B5: Custom URI schemes (18 evaluations, all gaps)}} \\
\addlinespace[1pt]
URI-01 & otpauth:/\allowbreak{}/\allowbreak{} (TOTP 2FA secret keys) & com.\allowbreak{}google.\allowbreak{}android.\allowbreak{}apps.\allowbreak{}authenticator2 /\allowbreak{} MainActivity & GAP & D & H \\
URI-02 & LPA:/\allowbreak{}/\allowbreak{}, com.android.euicc:/\allowbreak{}/\allowbreak{} (eSIM activation codes, carrier OAuth codes) & com.\allowbreak{}google.\allowbreak{}android.\allowbreak{}euicc /\allowbreak{} QrDownloadActivity, Redirect\-Uri\-Receiver\-Activity & GAP & D & H \\
URI-03 & FIDO:/\allowbreak{}/\allowbreak{} (FIDO2 authentication session data) & com.\allowbreak{}google.\allowbreak{}android.\allowbreak{}gms /\allowbreak{} QRBounceActivity & GAP & D & H \\
URI-04 & contactkeys:/\allowbreak{}/\allowbreak{} (E2E encryption key verification data) & com.\allowbreak{}google.\allowbreak{}android.\allowbreak{}contactkeys /\allowbreak{} MainActivity & GAP & D & H \\
URI-05 & tez:/\allowbreak{}/\allowbreak{}, gpay:/\allowbreak{}/\allowbreak{} (UPI payment data (payee VPA, amount)) & com.\allowbreak{}google.\allowbreak{}android.\allowbreak{}apps.\allowbreak{}nbu.\allowbreak{}paisa.\allowbreak{}user & GAP & D & H \\
URI-06 & genericidp:/\allowbreak{}/\allowbreak{}firebase.auth/\allowbreak{} (OAuth authorization codes (Firebase Auth SDK)) & 5 apps (Recorder, BetterBug, DeviceUsageStudy, Scribe, GMS) & GAP & D & H \\
URI-07 & gmail:/\allowbreak{}/\allowbreak{}label (Mail label names, search queries) & com.\allowbreak{}google.\allowbreak{}android.\allowbreak{}gm /\allowbreak{} Public\-Label\-Deep\-Link\-V2 & GAP & D & H \\
URI-08 & com.google.android.wear.companion:/\allowbreak{}/\allowbreak{}, comgooglecast:/\allowbreak{}/\allowbreak{}, vnd.youtube.kids:/\allowbreak{}/\allowbreak{} (OAuth authorization codes (5 endpoints)) & 3 apps (Wear Companion, Home, YouTube Kids) & GAP & D & H \\
URI-09 & pix:/\allowbreak{}/\allowbreak{} (Pix payment identifiers, amounts) & com.\allowbreak{}google.\allowbreak{}android.\allowbreak{}apps.\allowbreak{}walletnfcrel /\allowbreak{} DeepLinkActivity & GAP & D & H \\
URI-10 & comgooglewallet:/\allowbreak{}/\allowbreak{} (Card tokenization, Digital Car Key tokens) & com.\allowbreak{}google.\allowbreak{}android.\allowbreak{}gms /\allowbreak{} Add\-New\-Card\-Through\-Browser\-Activity, Entry\-Point\-Dck\-Activity & GAP & D & H \\
URI-11 & googlechrome:/\allowbreak{}/\allowbreak{} (URLs including auth tokens, session IDs) & com.\allowbreak{}android.\allowbreak{}chrome /\allowbreak{} IntentDispatcher & GAP & D & H \\
URI-12 & fitbit:/\allowbreak{}/\allowbreak{} (Health/\allowbreak{}fitness deep links, auth flow params) & com.\allowbreak{}fitbit.FitbitMobile & GAP & D & H \\
URI-13 & mpa:/\allowbreak{}/\allowbreak{} (Purchase order IDs, promo codes) & com.\allowbreak{}google.\allowbreak{}android.\allowbreak{}apps.\allowbreak{}tips /\allowbreak{} DeepLinkGateway & GAP & D & H \\
URI-14 & bbps:/\allowbreak{}/\allowbreak{}, bharatconnect:/\allowbreak{}/\allowbreak{}, upimeta:/\allowbreak{}/\allowbreak{} (Biller IDs, payment amounts, VPAs) & com.\allowbreak{}google.\allowbreak{}android.\allowbreak{}apps.\allowbreak{}nbu.\allowbreak{}paisa.\allowbreak{}user & GAP & D & H \\
URI-15 & searchlite:/\allowbreak{}/\allowbreak{}image/\allowbreak{} (Private search-result image URLs) & com.\allowbreak{}google.\allowbreak{}android.\allowbreak{}apps.\allowbreak{}searchlite & GAP & D & H \\
URI-16 & safetyhub:/\allowbreak{}/\allowbreak{}medicalinfo (Medical info deep link params) & com.\allowbreak{}google.\allowbreak{}android.\allowbreak{}apps.\allowbreak{}safetyhub & GAP & D & H \\
URI-17 & com.google.android.apps.youtube:/\allowbreak{}/\allowbreak{}oauth2redirect (OAuth authorization codes (missing state validation)) & com.\allowbreak{}google.\allowbreak{}android.\allowbreak{}youtube /\allowbreak{} Account\-Linking\-Activity & GAP & S+D & H \\
URI-18 & com.google.android.apps.gsa.gal:/\allowbreak{}/\allowbreak{}oauth2redirect (OAuth codes (shared library, all devices)) & com.\allowbreak{}google.\allowbreak{}android.\allowbreak{}googlequicksearchbox /\allowbreak{} Account\-Linking\-Activity & GAP & S+D & H \\
\addlinespace[3pt]
\multicolumn{6}{@{}l}{\textbf{B6: Platform security controls (10 evaluations, all gaps)}} \\
\addlinespace[1pt]
SYS-01 & ENFORCE\_\allowbreak{}STRICT\_\allowbreak{}QUERY\_\allowbreak{}BUILDER (ChangeId 143231523) globally disabled on Android 17 & All ContentProviders using this gate & GAP & S+D & H \\
SYS-02 & ENFORCE\_\allowbreak{}INTENTS\_\allowbreak{}TO\_\allowbreak{}MATCH\_\allowbreak{}INTENT\_\allowbreak{}FILTERS globally disabled on Android 17 & All Activities with intent filters & GAP & D & H \\
SYS-03 & BLOCK\_\allowbreak{}NULL\_\allowbreak{}ACTION\_\allowbreak{}INTENTS globally disabled on Android 17 & All intent receivers & GAP & D & H \\
SYS-04 & ENFORCE\_\allowbreak{}STRICT\_\allowbreak{}SQL\_\allowbreak{}CHECKS gated at targetSdk=37 (no Play Store app targets 37) & All ContentProviders & GAP & D & H \\
SYS-05 & PARCEL\_\allowbreak{}HARDENING gated at targetSdk=37 & All Parcelable IPC & GAP & D & H \\
SYS-06 & RESTRICT\_\allowbreak{}DATA\_\allowbreak{}URI\_\allowbreak{}COLUMNS gated at targetSdk=37 & ContentProvider URI grants & GAP & D & H \\
SYS-07 & DETECT\_\allowbreak{}IMPLICIT\_\allowbreak{}URI\_\allowbreak{}PERMISSION\_\allowbreak{}GRANT gated at targetSdk=37 & URI permission grants & GAP & D & H \\
SYS-08 & Settings Tile Parcelable write/\allowbreak{}read mismatch (CVE-2023-45777 regression) & Bundle-based IPC via AccountManager, notifications, activity results & GAP & S+D & C \\
SYS-09 & TLPE via InCallController create\-Package\-Context flag 3 (CVE-2026-49881) & system\_\allowbreak{}server code loading & GAP & S & X \\
SYS-10 & Maps AGMM JavaScript bridge openUrl() & JS-to-native intent redirect within Maps process & GAP & S & S \\
\addlinespace[3pt]
\multicolumn{6}{@{}l}{\textbf{Positive controls: well-defended components (11 evaluations, all holds)}} \\
\addlinespace[1pt]
NEG-01 & Pixel Launcher (all providers): Signature permissions, SHA-256 allowlists, test-mode gating & com.\allowbreak{}google.\allowbreak{}android.\allowbreak{}apps.\allowbreak{}nexuslauncher & HOLDS & S & P \\
NEG-02 & Google Meet: No intent forwarding pattern, no add\-Javascript\-Interface & com.\allowbreak{}google.\allowbreak{}android.\allowbreak{}apps.\allowbreak{}tachyon & HOLDS & S & P \\
NEG-03 & NFC stack (entire codebase): Flag 0 only, no FLAG\_\allowbreak{}MUTABLE, all Binder methods enforce permissions, user confirmation for dangerous NFC actions & com.\allowbreak{}android.\allowbreak{}nfc, com.\allowbreak{}google.\allowbreak{}android.\allowbreak{}nfc, com.\allowbreak{}google.\allowbreak{}android.\allowbreak{}pixelnfc & HOLDS & S & P \\
NEG-04 & Managed\-Provisioning: Code-level caller validation (verify\-Action\-And\-Caller, getCallingPackage, check\-Device\-Policy\-Preconditions, FRP check) compensates for missing manifest permission & com.\allowbreak{}android.\allowbreak{}managedprovisioning & HOLDS & S & P \\
NEG-05 & com.android.phone: All providers require permissions, dangerous RadioInfo operations gated by mSystemUser and Build.isDebuggable() & com.\allowbreak{}android.\allowbreak{}phone & HOLDS & S & P \\
NEG-06 & Gmail providers (10 tested): SecurityException, empty results, or not exported & com.\allowbreak{}google.\allowbreak{}android.\allowbreak{}gm & HOLDS & D & P \\
NEG-07 & GMS ContentProviders: Not exported or require permissions & com.\allowbreak{}google.\allowbreak{}android.\allowbreak{}gms & HOLDS & D & P \\
NEG-08 & FileProviders (Gmail, Wallet, Messages): Confirmed not exported & Multiple & HOLDS & S+D & P \\
NEG-09 & TLPE variant hunt (Maps, Camera, Search, Recorder): All resolve to Google-signed packages (Android Auto SDK, VR Core, GMS dynamite); all in app context, not system\_\allowbreak{}server & Multiple & HOLDS & S & P \\
NEG-10 & DynamiteModule pattern (\textasciitilde{}20+ apps): Hardcoded GMS package name & Multiple & HOLDS & S & P \\
NEG-11 & 17 ``dead ends'' in Vulnerability Reward Program testing: Components properly secured or not exploitable & Multiple (16 apps) & HOLDS & D & P \\
\end{longtable}
}

\end{document}